\documentclass[journal ]{AIAA-new}
\usepackage{amsmath}
\usepackage{bm}
\usepackage[utf8]{inputenc}
\usepackage{textcomp}

\usepackage{graphicx}
\usepackage{amsmath}
\newtheorem{definition}{Definition}

\newtheorem{lemma}{Lemma}
\usepackage{comment}
\usepackage[version=4]{mhchem}
\usepackage{siunitx}
\usepackage{longtable,tabularx}
\usepackage{multirow}
\usepackage{matlab-prettifier}

\title{Cislunar Mean-Motion Resonances: Definitions, Widths, and Comparisons with Resonant Satellites}

\author{Anjali Rawat \footnote{PhD Student, Aerospace and Ocean Engineering, Virginia Tech, Blacksburg, VA,  United States, anjalirawat@vt.edu}}
\affil{PhD Student, Aerospace and Ocean Engineering, Virginia Tech, Blacksburg, VA, 24061}
\author{Bhanu Kumar \footnote{James Van Loo Postdoctoral Assistant Professor, Department of Mathematics, University of Michigan, Ann Arbor, MI, United States, bhkumar@umich.edu}\footnote{NSF Postdoctoral Research Fellow, Jet Propulsion Laboratory, California Institute of Technology, Pasadena, CA, United States}}
\affil{Postdoctoral Assistant Professor, Department of Mathematics, University of Michigan, Ann Arbor, MI, 48109}
\affil{NSF Postdoctoral Research Fellow, Jet Propulsion Laboratory, California Institute of Technology, Pasadena, CA, 91109}
\author{Aaron J. Rosengren \footnote{Assistant Professor, Mechanical and Aerospace Engineering, University of California San Diego, La Jolla, CA, United States,  arosengren@ucsd.edu}}
\affil{Assistant Professor, Mechanical and Aerospace Engineering, University of California San Diego, La Jolla, CA, 92093}
\author{Shane D. Ross \footnote{Professor, Aerospace and Ocean Engineering, Virginia Tech, Blacksburg, VA, United States, sdross@vt.edu}}
\affil{Professor, Aerospace and Ocean Engineering, Virginia Tech, Blacksburg, VA, 24061}

\begin{document}

\maketitle

\begin{abstract}
Lunar mean-motion resonances (MMRs) significantly shape cislunar dynamics beyond GEO, forming stable-unstable orbit pairs with corresponding intermingled chaotic and regular regions. The resonance zone is rigorously defined using the separatrix of unstable resonant periodic orbits surrounding stable quasi-periodic regions. Our study leverages the planar, circular, restricted three-body problem (PCR3BP) to estimate the (stable) resonance widths and (unstable) chaotic resonance zones of influence of the 2:1 and 3:1 MMRs across various Jacobi constants, employing a Poincar\'e map at perigee and presenting findings in easily interpretable geocentric orbital elements. An analysis of the semi-major axis versus eccentricity plane reveals broader regions of resonance influence than those predicted by semi-analytical models based on the perturbed Kepler problem. A comparison with high-fidelity 3-dimensional ephemeris propagation of several spacecraft --- TESS, IBEX and Spektr-R --- in these regions is made, which shows good agreement with the simplified CR3BP model. 
\end{abstract}

\section*{Nomenclature}

{\renewcommand\arraystretch{1.0}
\noindent\begin{longtable*}{@{}l @{\quad=\quad} l@{}}
$C$ & Jacobi constant \\
$G$ & gravitational constant (km$^3$/kg$\cdot$s$^2$) \\
$H$, $h$, $G$, $g$, $L$, $\ell$  & Delaunay variables \\
$T$ & inertial-frame period of spacecraft (NDU) \\
$T_m$ & sidereal period of Moon (NDU) \\
$a$ & semi-major axis (NDU) \\
$a_m$ & semi-major axis of the Moon (km) \\
$e$ & eccentricity \\
$e_m$ & eccentricity of the Moon \\
$i$ & inclination w.r.t. Moon's orbital plane (rad) \\
$k$ & number of geocentric orbits completed by spacecraft in Earth-centered inertial frame \\
$k_m$ & number of geocentric orbits completed by Moon in Earth-centered inertial frame \\
$m_e$, $m_m$ & mass of Earth, Moon (kg) \\
$m_1$, $m_2$ & mass of primary, secondary (NDU) \\
$n_m$ & Moon's mean motion ($s^{-1}$)\\
$r_1$ & distance from spacecraft to primary (NDU) \\
$r_2$ & distance of spacecraft from secondary (NDU) \\
$\ell$ & mean anomaly (rad) \\
$\varepsilon$ & obliquity of the ecliptic (rad) \\
$\mu$ & mass ratio of Earth-Moon system (NDU) \\
$\mu_e$, $\mu_m$ & gravitational parameter of Earth ($Gm_e$), Moon ($Gm_m$) (km$^3$/s$^2$) \\
$\nu$ & true anomaly (rad) \\
$\lambda$ & spacecraft's mean longitude (rad) \\
$\lambda_m$ & Moon's mean longitude (rad) \\
$\omega$ & argument of perigee (rad) \\
$\omega_m$ & angular velocity of Moon (rad/s) \\
$\Omega$ & longitude of ascending node in synodic frame (rad) \\
$\Omega_{iner}$ & longitude of ascending node in inertial frame (rad) \\
$\varpi$ & longitude of perigee in synodic frame(rad) \\
$\varpi_{iner}$ & longitude of perigee in inertial frame (rad) \\
$\mathbf{r}$ & position vector of spacecraft in Earth-centered inertial frame (km) \\
$\mathbf{v}$ & velocity vector of spacecraft in Earth-centered inertial frame (km/s) \\
$\mathbf{r}_m$ & position vector of Moon in Earth-centered inertial frame (km) \\
$\mathbf{v}_m$ & velocity vector of Moon in Earth-centered inertial frame (km/s) \\
$\mathbf{v}_{m,cir}$ & velocity of the Moon under circular motion assumption (km/s) \\
$\mathbf{r}_{syn}$ & position vector of spacecraft in synodic frame (km) \\
$\mathbf{v}_{syn}$ & velocity vector of spacecraft in synodic frame (km/s) \\
$\mathbf{r}_{syn, non-dim}$ & non-dimensionalized position vector of spacecraft in synodic frame (NDU) \\
$\mathbf{v}_{syn, non-dim}$ & non-dimensionalized velocity vector of spacecraft in synodic frame (NDU) \\
$\mathbf{\hat{x}}$, $\mathbf{\hat{y}}$, $\mathbf{\hat{z}}$ & instantaneous unit vectors for synodic frame \\
$\dot{\hat{\mathbf{x}}}$, $\dot{\hat{\mathbf{y}}}$, $\dot{\hat{\mathbf{z}}}$ & instantaneous rate of change of unit vectors for synodic frame \\
$R_{iner}^{syn}$ & rotation matrix to convert from geocentric inertial ecliptic to synodic frame\\
$R_{eci}^{eclip}$ & rotation matrix to convert from geocentric equatorial to ecliptic frame\\

\multicolumn{2}{@{}l}{Subscripts}\\
$e$	& Earth \\
$m$ & Moon\\
\end{longtable*}}

\section{Introduction}
\label{sec:intro}

The dynamics of cislunar space beyond geosynchronous orbit (xGEO) are fundamentally influenced by mean-motion resonances (MMRs), a phenomenon previously underappreciated in Earth-satellite dynamics 
due to their negligible effect on the traditional geocentric domains.
However, for missions like IBEX and TESS, which operate within predominant lunar MMRs, 
and others like Spektr-R, which seemingly navigates unstable resonance regions, recognizing the impact of MMRs is crucial. 
Determining the stability of a space asset's xGEO orbit necessitates a thorough understanding of the dynamical structure of MMRs, particularly the extent of stable MMRs and the surrounding chaotic regions across various semi-major axis values.

Semi-analytical methods, such as Gallardo's algorithm \cite{gallardo2006atlas}, have been used to assess the domains of influence of predominant MMRs in planetary dynamics. Yet, these methods often presuppose constant eccentricity over
resonant timescales, which does not accurately reflect the highly perturbed Earth-Moon environment of xGEO. Such approaches, moreover, are fundamentally based on the perturbed-Hamiltonian formulation, which provides a local, rather than global, description of phase space; they consequently underestimate the true regions of influence of MMRs. A global geometric dynamical portrait can be furnished by semi-analytical approaches to the circular, restricted three-body problem (CR3BP). Although some methods incorporate the full CR3BP model, they rely on narrowly defined initial conditions that generate localized Poincar\'e maps specific to a given resonance \cite{malhotra2020divergence}, thus overlooking the intricate global dynamics of the phase space. While \cite{winter1997resonance} employs a Poincar\'e map at perigee to determine the precise widths of interior first-order resonances in the Sun-Jupiter system, their analysis does not include visualizations of these maps in orbital elements, which offer more interpretability. Additionally, all of these studies tend to neglect the presence of unstable resonant periodic orbits and quantification of the chaotic resonance zones, which are critical for understanding transit through large connected chaotic regions that dominate for a large range of Jacobi-constant values.

This study computes the semi-major axis {\it stable resonance widths} and the larger circumscribing {\it unstable resonance zones} of influence for key lunar MMRs, specifically the 2:1 and 3:1 resonances, using the planar (P)CR3BP and following known methods for its study \cite{Meiss1992,KoLoMaRo2000,RoSc2007,NaikLekienRoss2017,hiraiwa2024designing}. 
By applying a Poincar\'e map at the perigee of osculating orbits \cite{RoSc2007,hiraiwa2024designing},
we delineate these regions in the plane of semi-major axis and {\it synodic} longitude of perigee $\varpi$ (the longitude of perigee relative to the Earth-Moon line). Our Poincar\'e maps reveal resonance regions, notably the prominent 2:1, 3:1, and 4:1 resonance ``islands'', through which we can determine the {\it stable} width of a resonance, defined as the semi-major axis span of the ``largest'' (i.e., outermost) stable quasi-periodic torus. 
According to the Poincar\'e-Birkhoff theorem \cite{poincare1912theorem,birkhoff1913proof}, stable resonant points (the centers of the islands) will alternate with unstable resonant points, which will be embedded in a (perhaps large) chaotic set.
These unstable resonant periodic orbits can be computed via symmetry, and their stable and unstable manifolds calculated and  visualized on the same Poincar\'e map. 
Chaotic (resonance) zones are identified as regions enclosed by segments of 
stable and unstable manifolds that form the regions' boundaries, according to well-established dynamical-systems methods \cite{MaMePe1984,wiggins1990geometry}.
Resonance widths and the larger enclosing chaotic resonance zones are computed across a range of Jacobi constants, subsequently correlating the widths with projections of PCR3BP energy surfaces onto the osculating eccentricity ($e$) versus semi-major axis ($a$) plane. 
The directly computed PCR3BP-based stable resonance zone widths are compared with semi-analytical predictions \cite{gallardo2006atlas,tG19,tG20}.
Moreover, the orbits of current xGEO spacecraft like TESS, IBEX and Spektr-R, obtained via their two-line element (TLE) sets, are projected onto the $(a,e)$-plane to ascertain their positioning  
within stable or unstable resonance regions. The perigee mappings derived from higher-fidelity ephemeris propagations of TESS, IBEX and Spektr-R are superimposed onto the Poincar\'e map and $(a,e)$-plane, further validating the PCR3BP stable resonance widths and chaotic resonance zones.

This paper is structured into nine sections. Section~\ref{sec:intro} introduces the problem's motivation and reviews relevant prior research. Section~\ref{sec:CR3BP} provides a concise overview of the CR3BP. Section~\ref{sec:poincare} details key aspects of Poincar\'e maps, focusing on stable and unstable periodic orbits, and defining stable resonance widths and the larger chaotic resonance regions.
Section~\ref{sec:resonances} explores MMRs in xGEO, highlighting both stable and unstable periodic orbits. Section~\ref{sec:manifolds} outlines the methodology for computing unstable resonant orbits and their stable and unstable manifolds. Section~\ref{sec:gallardo} presents summarized findings from a semi-analytical approach assessing MMR widths. Section~\ref{sec:zones} discusses the methodology employed to determine resonance widths and chaotic zones, mapping them onto the $(a,e)$ plane, and compares with semi-analytically computed widths and TLEs of TESS, IBEX and Spektr-R. Section~\ref{sec:ephem} explores the high-fidelity trajectories of TESS, IBEX, and Spektr-R using JPL Horizons ephemeris data and a Cowell 4-body ephemeris propagation and maps them onto $(a,e)$ and $(\varpi,a)$ planes. Finally, Section~\ref{sec:conclusion} provides a  summary of the paper's results and future work.

\section{Planar, Circular, Restricted Three-Body Problem}
\label{sec:CR3BP}

The PCR3BP is the simplest model for motion in cislunar space whose dynamics capture the main qualitative features of the true motion. 
It describes the motion of a massless spacecraft relative to two primary bodies (e.g., Earth and Moon), viewed in a rotating reference frame centered at the center of mass (barycenter) of the two primaries. The PCR3BP assumes that both primaries  move in circular orbits with constant angular velocity about their barycenter, and that the spacecraft motion is coplanar with their orbits.

\begin{figure}[hbt!]
	\centering
    \begin{tabular}{cc}
	\includegraphics[width=0.4\linewidth]{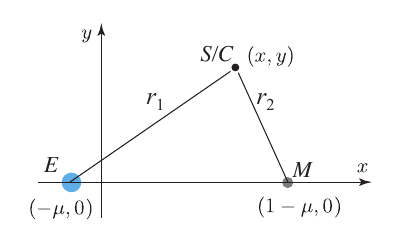} &
    \includegraphics[width=0.4\linewidth]{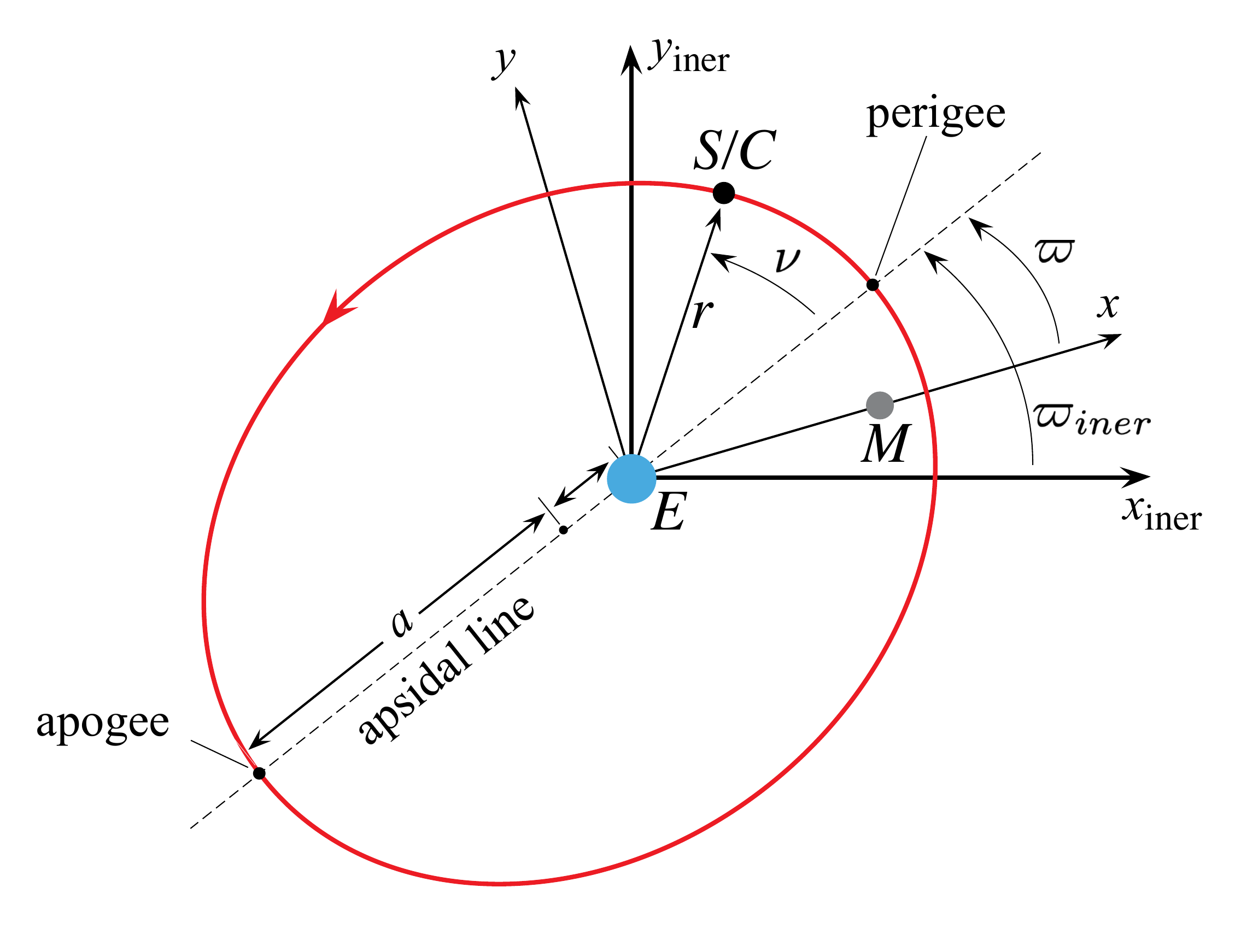}\\ 
        (a)  & (b) 
     \end{tabular}

    \caption{
     (a) Non-dimensional barycentered co-rotating $(x,y)$ frame.
     (b) The geocentric osculating orbital elements showing inertial longitude of perigee ($\varpi_{iner}$) and synodic longitude of perigee ($\varpi$).
    }   
	\label{frame}
\end{figure}

In the equations of motion of the PCR3BP, we choose normalized units such that the distance between the two masses $m_1$ and $m_2$ is 1, their combined mass is 1, and the period of their orbit relative to the barycentered inertial frame, i.e., the sidereal period $T_m$, is $2\pi$. 
The only parameter of the system dynamics is then the mass ratio $\mu$, defined as $\frac{m_2}{m_1+m_2}$. 
For the Earth-Moon system, we use 
$\mu =  1.2150584270571545 \times10^{-2}$.
We choose a coordinate frame rotating with the two massive bodies as in Fig.~\ref{frame}(a), centered at their barycenter with $m_1$ and $m_2$ lying on the $x$-axis at $(-\mu,0)$ and $(1-\mu,0)$, respectively. The PCR3BP second-order differential equations of motion for the spacecraft in normalized units are then

\begin{equation}
\begin{split}
\ddot{x} -2\dot{y}&=x -(1-\mu)\frac{x+\mu}{r_{1}^{3}}-\mu \frac{x-1+\mu}{r_{2}^{3}}, \\
\ddot{y} +2\dot{x}&=y-(1-\mu)\frac{y}{r_{1}^{3}}-\mu \frac{y}{r_{2}^{3}},
\label{PCR3BP} 
\end{split}
\end{equation}
where $r_{1} = \sqrt{(x+\mu)^{2} + y^{2}}$ is the distance from the spacecraft to $m_1$ and $r_{2} = \sqrt{(x-1+\mu)^{2} + y^{2}} $ is the distance to 
$m_2$. In general, we refer to a point in the 4-dimensional, phase-space manifold $\mathcal{M}$ as $X$. This point can be written in terms of the rotating-frame Cartesian coordinates given above, $X=(x,y,\dot x,\dot y)$. 
Alternatively, in the geocentric part of $\mathcal{M}$, one can use instantaneous (i.e., osculating), geocentric orbital elements, e.g., 
$X = (a,e,\ell,\varpi)$ where $a$ is the semi-major axis, $e$ the eccentricity, $\ell$ the mean anomaly, and $\varpi$ the longitude of perigee with respect to the rotating frame positive $x$-axis, as depicted in Fig.~\ref{frame}(b).

\paragraph{Jacobi Constant and the Energy Manifold.}
The Jacobi constant is proportional to the negative of the Hamiltonian energy of the system and is a constant of motion of the CR3BP equations. 
In other words, for an initial condition $X\in\mathcal{M}$, this scalar value does not change along the trajectory. The formula we use for the Jacobi constant of the planar problem is \footnote{We note that this definition differs from some authors, who add a constant value $\mu(1-\mu)$, so that the Jacobi constant of the $L_4$, $L_5$ points is precisely 3. We adopt the convention in current use among the cislunar astrodynamics community.}
\begin{equation} 
\label{jacobi} \mathcal{C}(x,y,\dot x,\dot y) = x^{2} + y^{2}  + 2\left( \frac{1-\mu}{r_{1}}+ \frac{\mu}{r_{2}} \right) - \left( \dot x^{2} + \dot y^{2}  \right). 
\end{equation}
Let $\mathcal{M}_C$ be the  energy manifold or
energy surface given by setting the Jacobi integral \eqref{jacobi} equal to a constant, i.e., 

\begin{equation}
\mathcal{M}_C=\{X\in\mathcal{M} \mid \mathcal{C}(X)=C = {\rm constant}\}.
\end{equation}

The surface $\mathcal{M}_C$ can be considered as a 3-dimensional manifold embedded in the 4-dimensional phase space $\mathcal{M}$. Any PCR3BP trajectory starting in $\mathcal{M}_C$ will remain therein for all time, so one can study PCR3BP dynamics restricted to each 3D manifold $\mathcal{M}_C$ separately. 
For the geocentric portion of $\mathcal{M}_C$ interior to the Moon's orbit, dimensionality can be further reduced by using a 2-dimensional Poincar\'e surface of section, described in Section~\ref{sec:poincare}.

\paragraph{Tisserand Approximation.} 
The Tisserand approximation to the Jacobi constant in geocentric orbital-element space in the CR3BP model is
\begin{equation}
    C(a,e,i) = \frac{1}{a} + 2 \sqrt{a(1-e^2)}\cos{i} + {O}(\mu),
\end{equation}
which dynamically limits the range of motion a non-maneuvering spacecraft is capable of, depicted using the geocentric, osculating orbital elements: semi-major axis ($a$), eccentricity ($e$), and inclination ($i$). In the PCR3BP, the inclination is set to zero ($i=0$). The terms of order $\mu$, ${O}(\mu)$, will be ignored when the Tisserand approximation is used to plot projections of $\mathcal{M}_C$ onto $(a,e)$ space, which appear as curves, referred to as {\it Tisserand curves}. 

\section{The Poincar\'e Map and Key Dynamical Features} \label{sec:poincare}
\subsection{Surface of Section at Perigee}
In our study, we define the Poincar\'e surface of section at perigee crossings identified when the geocentric mean anomaly $\ell$ is zero (same condition as true anomaly $\nu$ equals zero). Our Poincar\'e section, parametrized by Jacobi constant $C$, is defined as,
\begin{equation}
    \Sigma_C = \{ X \in \mathcal{M}_C ~|~ \ell = 0 \}.
\end{equation}

In practice, to avoid numerical problems in detecting section crossings during propagation, we consider a continuous function of the mean anomaly which is zero at $\ell = 0$, as described in Appendix A. The Poincar\'e section so constructed is 2D, and can be represented by two coordinates that can be interpreted in orbital-element space: the semi-major axis $a$ and the synodic longitude of perigee $\varpi$, the angle between perigee and the Moon's location in the rotating frame (see Fig. \ref{frame}). As $\varpi$ is an angular variable, $\Sigma_C$ has a cylindrical topology, i.e., $(\varpi,a) \in S^1 \times I$ where $I \subset \mathbb{R}$, and $S^1$ is the circle. 

\subsection{Poincar\'e Map on the Poincar\'e Section}

Poincar\'e maps simplify the study of the PCR3BP by transforming a 4-dimensional phase space
into a more manageable 2-dimensional analysis, elucidating periodic, quasi-periodic, and chaotic behaviors, and revealing the

\begin{figure}[hbt!]
\centering
\includegraphics[scale = 0.4]{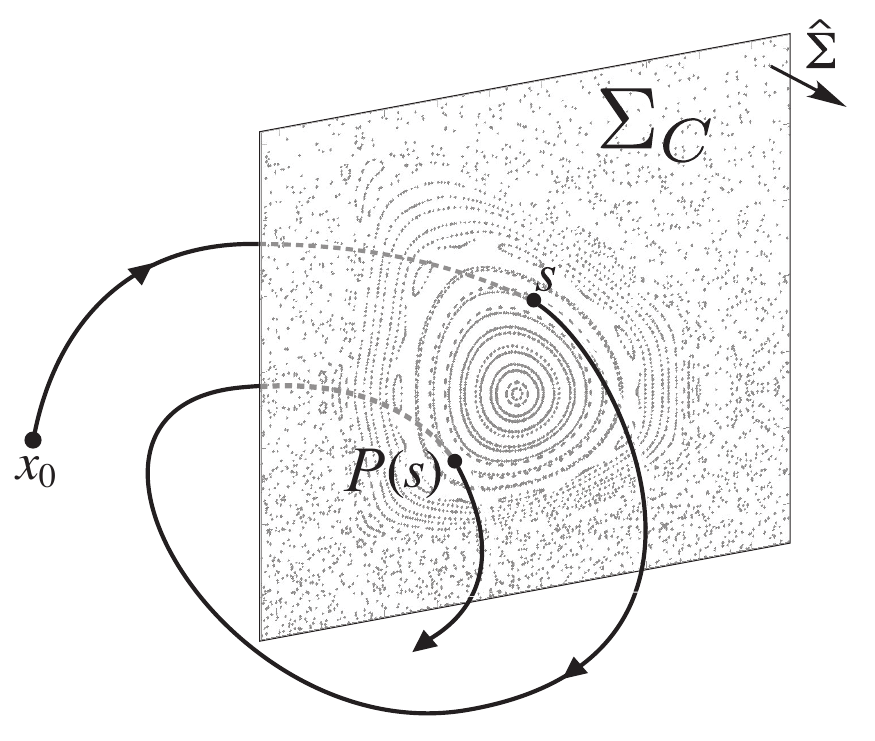}
\caption{\label{fig:PM}Poincar\'e map $P$ on a Poincar\'e section $\Sigma_C$ in the CR3BP. The unit vector $\hat{\Sigma}$ gives the sense in which trajectories are crossing $\Sigma_C$.}
\end{figure}
intricate manifold structures that govern the system's dynamics. The section $\Sigma_C$ represents a 2-dimensional surface transverse to the flow inside the energy manifold $\mathcal{M}_C$ (cf.\ Fig.~\ref{fig:PM}). Let $x_0 \in \mathcal{M}_C$ denote an initial state, not necessarily in $\Sigma_C$. The Poincar\'e map $P(x_0)=s$ corresponds to the first crossing of $\Sigma_C$ by the trajectory originating at $x_0$ in a particular direction. In general, we will consider Poincar\'e mappings of $\Sigma_C$ to itself,
\begin{equation}
    \begin{split}
    \label{poincare_map} 
    P : \Sigma_C  & \rightarrow \Sigma_C, \\
    s & \mapsto P(s).
    \end{split}  
\end{equation}
We establish important terminology below.

\begin{figure}[hbt!]
	\centering
    \begin{tabular}{cc}
	\includegraphics[width=0.47\linewidth]{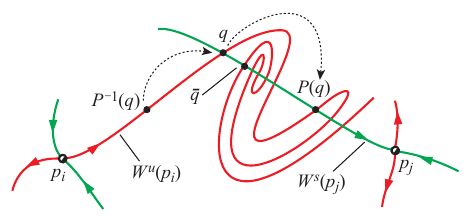} &
    \includegraphics[width=0.47\linewidth]{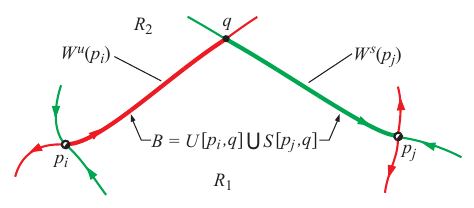}\\ 
        (a)  & (b) 
     \end{tabular}
	\caption{(a) PIP ($q$) and secondary intersection point  ($\bar q$, not a PIP). (b) A BIP $q$ defining a local boundary $B$ between two regions $R_1$ and  $R_2$.}
	\label{fig:pip}
\end{figure}

\begin{definition}
The ``orbit'' of a point $s \in \Sigma_C$ under $P$ is the set of all the past and future iterates of the point $s$ under the map $P$, i.e., the infinite sequence of points,
$\{ \ldots,P^{-1}(s),s, P^{1}(s),P^{2}(s), \ldots \}$, also denoted as $\mathcal{O}(s)$. 
See Figs.~\ref{fig:PM} and \ref{fig:pip}(a) for examples of points and their iterates under the map $P$.
Note that the orbit of $s$ is the same as the orbit of $P^k(s)$ for all $k\in \mathbb{Z}$, and all represent the same continuous ``trajectory'' within the energy manifold $\mathcal{M}_C$.
\end{definition}
Some orbits of $P$ do not contain an infinite number of distinct points, but instead have a finite number of distinct points. These are periodic orbits of $P$ in the sense that the sequence repeats after some minimum integer $n\ge1$ number of iterates.

\begin{definition}
A periodic orbit of $P$ is a finite sequence of points
$\mathcal{O}(p_n) = \{p_1, \ldots, p_{n}\}$ such that $p_k = P(p_{k-1})$ for $2\le k \le n$ and $p_1= P(p_n)$.
The period-$n$ points $\mathcal{O}(p_n)$ represent a continuous periodic PCR3BP trajectory, a closed loop, within the energy manifold $\mathcal{M}_C$.
\end{definition}
\vspace{-2mm}
This is a generalization of fixed points, as the state returns to the initial point $p_1\in\Sigma$ after $n$ iterates of the Poincar\'e map, i.e., $p_1=P^n(p_1)$, where $P^n$ denotes $n$ compositions of $P$, $P\circ P \circ \cdots \circ P \circ P$ ($n$ times). 
For a periodic orbit, we note that each of the points $p_1,\ldots,p_n$ is a fixed point (a period-1 point) under $n$ iterates of the map, i.e., $p_k = \bar P (p_k)$, where $\bar P = P^n$, and thus we may occasionally refer to each one individually as a fixed point.

If a periodic orbit $\mathcal{O}(p_n)$ is of saddle-type, each of the points $p_1,\ldots,p_n \in \mathcal{O}(p_n)$ will have attached stable $(W^s(p_i))$ and unstable $(W^u(p_i))$ invariant manifolds, for $i = 1, \ldots, n$, 
consisting of orbits of $P$ which tend asymptotically toward  $\mathcal{O}(p_n)$ forwards or backwards in time, respectively.
It is known that for 2D Poincar\'e maps of 2 degree-of-freedom Hamiltonian systems \cite{MeOt1986,ScOt1997}, some periodic orbits $\mathcal{O}(p_n)$ are related to a general idea of ``resonance'', even beyond just the application to orbital dynamics. 
If $\mathcal{O}(p_n)$ is of center-type stability, this is a stable resonant orbit. 
If $\mathcal{O}(p_n)$ is of saddle-type stability, this is an unstable resonant orbit. It should be noted that all members $p_1,\ldots,p_n$ of a periodic orbit have the same stability type.

Let us focus for now on an unstable resonant orbit. There is a systematic way to obtain a ``resonance zone'' --- or, as mentioned earlier, the chaotic resonance zone corresponding to the resonance --- via the stable and unstable manifolds of  $\mathcal{O}(p_n)$.
We first must define a certain type of intersection between the stable and unstable manifolds of  $\mathcal{O}(p_n)$.
As the stable and unstable manifolds $W^s(p_i)$ and $W^u(p_j)$ are 1-dimensional curves within the 2-dimensional Poincar\'e section $\Sigma_C$, they will generally intersect transversally in points.\footnote{A degenerate case occurs when $W^s(p_i)=W^u(p_j)$, that is, a 1-dimensional intersection---an unbroken separatrix---but this case is not considered.}
An intersection point of the stable and unstable manifolds of the saddle-type periodic orbit 
$\mathcal{O}(p_n)=\{p_1, \ldots, p_{n}\}$ is termed a primary intersection point (PIP), denoted by $q$ in Fig.~\ref{fig:pip}(a), if it meets the following criteria. 

\begin{definition}
Suppose $q \in W^u(p_i) \bigcap W^s(p_j)$, where $p_i,p_j \in \mathcal{O}(p_n)$,
and let $U[p_i,q]$ denote the segment of $W^u(p_i)$ with endpoints $p_i$ and $q$ and  $S[p_j,q]$ denote the segment of $W^s(p_j)$ with endpoints $p_j$ and $q$.  Then $q$ is called a  ``primary intersection point'' (PIP) if $U[p_i,q]$ intersects $S[p_j,q]$ only at the point $q$ (and $p_i$ if $i=j$).
\end{definition}
Since PIPs are intersections of stable and unstable invariant manifolds, it follows from the definition of stable/unstable invariant manifolds, that all the past and future iterates of a PIP are also PIPs. The following lemma is proved in Wiggins 1990 \cite{wiggins1990geometry}.

\begin{lemma}
Suppose $q \in W^u(p_i) \bigcap W^s(p_j)$ is a PIP; then $P^k(q)$ is a PIP for all $k \in \mathbb{Z}$.
\end{lemma}

\noindent We use PIPs to define {\it boundaries}, sometimes called separatrices, and subsequently resonance regions.
Any PIP can be used to denote a local boundary. All intersections between stable and unstable invariant manifolds of the same periodic orbit mark homoclinic points as they enable homoclinic transfers. For example, $q$ and $\bar{q}$ are a homoclinic points for periodic orbit $\mathcal{O}$($p_n$) in Fig.\ref{fig:pip}.

\begin{definition}
Suppose $W^u(p_i)$ and $W^s(p_j)$ intersect in the PIP $q$.
Define $B \equiv U[p_i,q] \bigcup S[p_j,q]$ as
a boundary between two ``sides,'' region $R_1$ and region $R_2$.
The PIP is then called a boundary intersection point (BIP).
\end{definition}
As a matter of convention, for a BIP we pick the
PIP with the shortest arc-length of the manifolds, measured from the fixed points to the intersection point, i.e., the shortest arc-length for $U[p_i,q] \bigcup S[p_j,q]$. 
The BIP and the boundary $B$ it defines  allows for the local division of the Poincar\'e section $\Sigma_C$ into distinct regions $R_1$ and $R_2$, as illustrated in
Fig. \ref{fig:pip}(b). 

\begin{figure}[hbt!]
	\centering
    \includegraphics[width=0.45\linewidth]{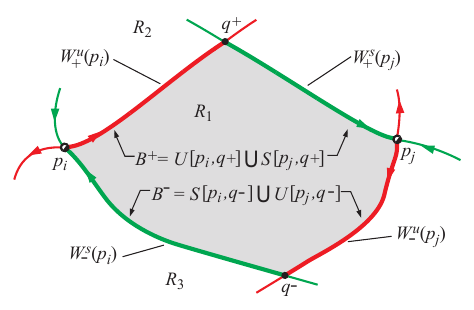} 
	\caption{Construction of a top and bottom boundary to a resonance region $R_1$.
 }
	\label{fig:boundary_both}
\end{figure}

As there are always two branches each for a stable and unstable manifold, we can identify both a ``top'' and ``bottom'' boundary.
We denote the two branches of the unstable manifold of a period-$n$ point $p_i$ by $W^u_+(p_i)$ and $W^u_-(p_i)$, and similarly for $p_j$.  
Referring to Fig.~\ref{fig:boundary_both}, we suppose that $W^u_+(p_i)$ intersects 
$W^s_+(p_j)$ and $W^s_-(p_i)$ intersects $W^u_-(p_j)$.

 Choosing BIPs $q^+ \in W^u_+(p_i) \bigcap W^s_+(p_j)$ and $q^- \in W^s_-(p_i) \bigcap W^u_-(p_j)$, we can identify boundaries $B^+$ and $B^-$ that define a closed region $R_1$, which we refer to as a {\it resonance region}. 

\begin{figure}[hbt!]
	\centering
    \includegraphics[width=0.7\linewidth]{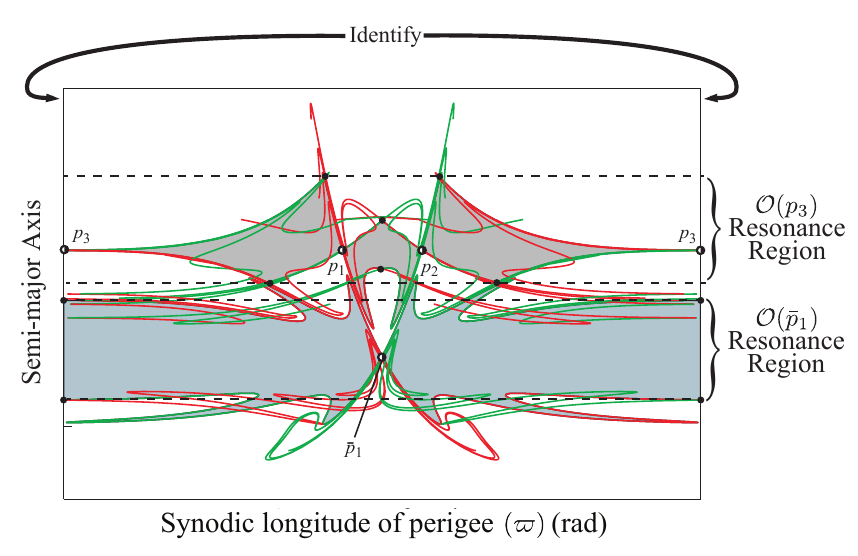} 
	\caption{Resonance regions for a period-3 orbit $\mathcal{O}(p_3)=\{p_1,p_2,p_3\}$ and a neighbouring period-1 orbit $\mathcal{O}(\bar p_1)=\{\bar p_1 \}$. The ``identify'' indicates that the synodic longitude of perigee, is an angle.
    }
	\label{fig:two-resonance-regions}
\end{figure}

Within our cylindrical Poincar\'e section $\Sigma_C$, we can have multiple resonance regions. Fig. \ref{fig:two-resonance-regions} shows an example calculation \cite{KoLoMaRo2022} of two neighboring resonance regions corresponding to neighboring resonant orbits.
For our purposes, we will label the ``width'' of a (chaotic) resonance region containing unstable resonant  orbit $\mathcal{O}(p_n)$ as the distance between the BIPs of $\mathcal{O}(p_n)$ having maximum and minimum semi-major axes (the dashed lines in Fig. \ref{fig:two-resonance-regions}).

\begin{figure}[hbt!]
	\centering
    \includegraphics[width=0.82\linewidth]{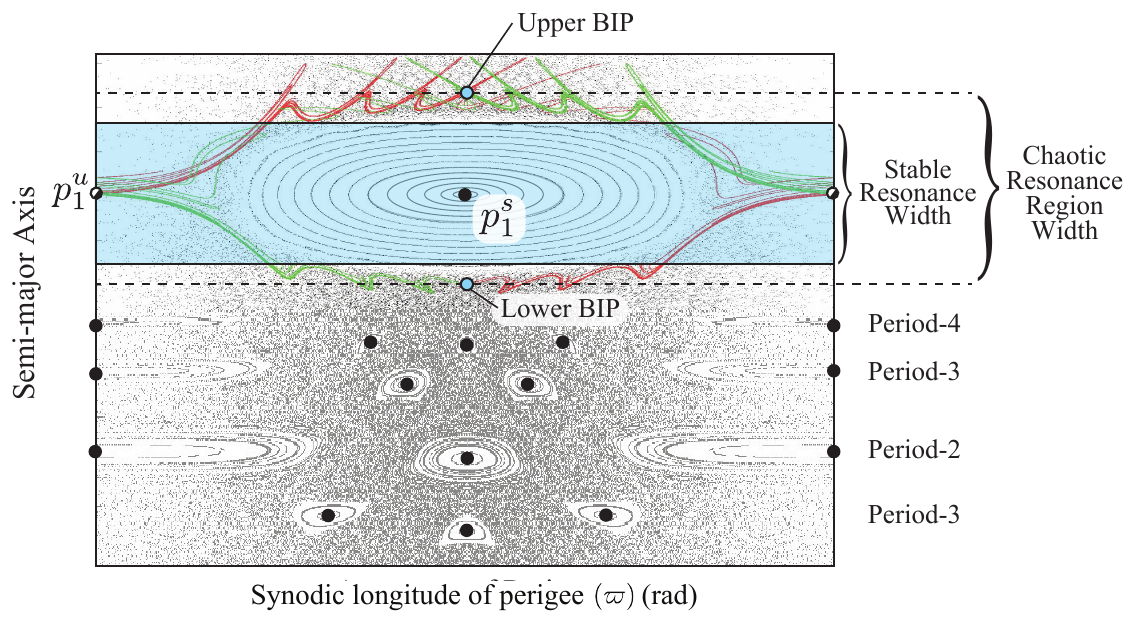} 
	\caption{Stable resonance widths (corresponding to stable period-1 resonant orbit $p_1^s$) and chaotic resonance regions (corresponding to its unstable counterpart $p_1^u$). Other stable period-$n$ points are also shown.  
 }
	\label{fig:resonance_zone_stable_and_bips}
\end{figure}

Resonance regions represent the dynamical ``sphere of influence'' of a particular resonance. In the PCR3BP, they contain within them the corresponding stable resonant periodic orbit and the surrounding stable quasi-periodic orbits, the librational resonant tori. Outside of the largest (outermost) librational torus, there is a ``stochastic'' or chaotic layer. The stable resonance width is given by the outermost closed curve of the stable resonance, while the larger chaotic resonant region width is given by the semi-major axis width between the upper and lower BIP.
This is illustrated in Figure \ref{fig:resonance_zone_stable_and_bips} which shows an example calculation on a Poincar\'e section at perigee, $\Sigma_C$, showing background points; initial conditions followed over several iterates of the map $P$.
The dynamics of motion into and out of the resonance region are determined by lobe dynamics not addressed here \cite{KoLoMaRo2022}, but documented elsewhere \cite{RoWi1990,KoMaRoLoSc2004,DeJuKoLeLoMaPaPrRoTh2005,RoTa2012,NaikLekienRoss2017}.

Intersections between stable and unstable invariant manifolds of different periodic orbits can occur. 
For example, suppose $\mathcal{O}(p_n)$ and $\mathcal{O}(\bar p_{\bar{n}})$  
correspond to two different continuous unstable periodic trajectories in the PCR3BP, as in Fig. \ref{fig:two-resonance-regions}. Unless constrained by other barriers within $\Sigma_C$ such as rotational invariant curves (RICs) \cite{Meiss1992,RoSc2007,werner2022multiple} ---  quasiperiodic KAM tori that block transport along semi-major axis in the cylindrical phase space of $\Sigma_C$ --- it is possible for there to be intersections between stable and unstable manifolds of $\mathcal{O}(p_n)$ and $\mathcal{O}(\bar p_{\bar{n}})$. Such intersections are denoted as {\it heteroclinic} points and enable heteroclinic transfers. Although not explicitly labeled, one can observe several  heteroclinic points in Fig.~\ref{fig:two-resonance-regions} between the displayed unstable resonant orbits $\mathcal{O}{(p_3)}$ and $\mathcal{O}{(\bar p_1)}$. Heteroclinic transfers between unstable resonant orbits will be explored in future work.

\section{Resonance in xGEO}
\label{sec:resonances}

\begin{figure}[hbt!]
	\centering
    \includegraphics[width=0.65\linewidth]{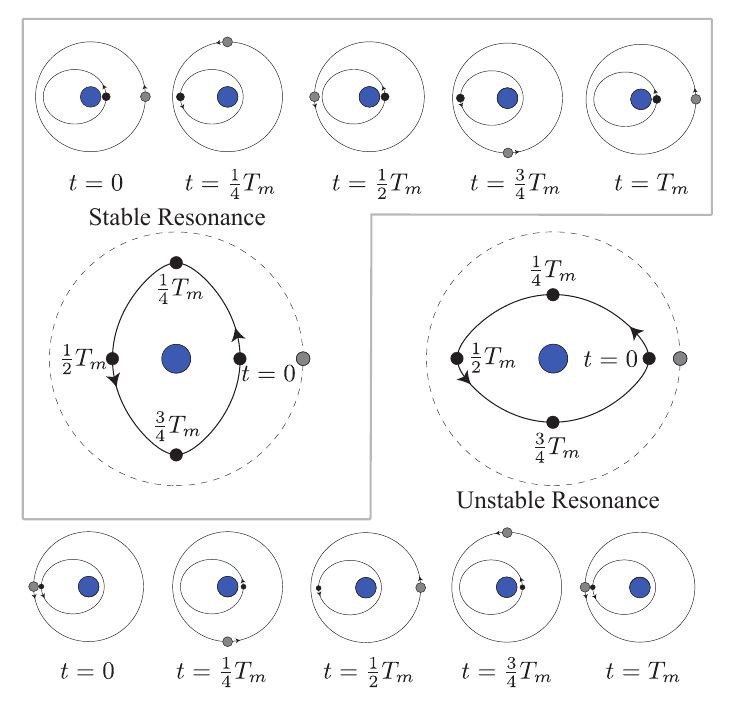} 
	\caption{(Top left) Configurations of 2:1 stable mean-motion resonance at fractions of $T_m$ (Moon's orbital period).
 (Bottom right) Configurations of 2:1 unstable mean-motion resonance at fractions of $T_m$.
 }
	\label{fig:stable-and-unstable-resonances-schematic}
\end{figure}

Mean-motion resonances in celestial mechanics result from gravitational interactions among celestial bodies, where their orbital periods create specific integer ratios. For instance, a 2:1 resonance signifies that one body completes two orbits for every single orbit completed by another, thereby influencing the stability and evolutionary trajectories of orbits over extended duration; see Fig.~\ref{fig:stable-and-unstable-resonances-schematic}.
Of particular interest for space domain awareness (SDA) within cislunar xGEO space are recurrent pathways between Earth and the Moon's orbit. Resonant orbits, being oftentimes periodic or quasi-periodic, have a rich history in mission design applications, particularly within the Earth-Moon system. 

A mean-motion resonance, denoted as $k$:$k_m$, is characterized by the ratio of orbital periods, where $k$ and $k_m$ are coprime positive integers respectively representing the number of spacecraft and Moon orbits completed around the Earth in equal time in an Earth-centered inertial frame. In this paper, we study interior resonances, where $k > k_m$.
Expressed in terms of the inertial-frame period of the spacecraft, $T$, and the sidereal period of the Moon, we have an approximate relationship, $T/T_m \approx k_m/k$. When observed within the context of the CR3BP, resonant orbits do not adhere strictly to the integer ratio $k_m/k$; thus
$T/T_m \approx k_m/k$ is approximate. Instead, a spacecraft completes approximately $k$ revolutions around the Earth in the time it takes the Moon to complete $k_m$ revolutions. 
In terms of the Poincar\'e map and section described above, $(P,\Sigma_C)$, a resonant orbit is a period-$k$ orbit of $P$ which takes approximately $k_m$ sidereal lunar periods of physical time.
These orbits can be found via differential-correction procedures that are well-known in the literature \cite{KoLoMaRo2022}.

Resonant orbits can be categorized as either stable (characterized for interior resonances by a perigee oriented towards the Moon viewed in the rotating frame) or unstable (for interior resonances, having an apogee oriented towards the Moon) as shown in Fig.\ref{fig:stable-and-unstable-resonances-schematic}. 
Unstable resonant orbits can be used in mission design for transfer scenarios, while stable resonant orbits, as demonstrated by missions such as IBEX and TESS, ensure sustained operational stability. For example, IBEX transitioned into a stable 3:1 resonant orbit following its launch, contributing to its prolonged mission duration \cite{Dichmann2013}. Similarly, TESS has maintained a stable 2:1 resonant orbit since its inception via a lunar flyby \cite{Ricker2014}.

\section{Unstable Periodic Orbits and Manifolds}
\label{sec:manifolds}

In the PCR3BP, to understand the structure of unstable resonant orbit families and the heteroclinic dynamics induced by them, one needs to compute the corresponding periodic orbits as well as their stable/unstable manifolds. To compute a family of $k$:$k_m$ unstable periodic orbits in the Earth-Moon PCR3BP, we start with an orbit state from the Earth Kepler problem having semi-major axis $a$ such that $\left(a/a_m\right)^{3/2}=\left(k_m/k\right)$, where $a_m=1$ in non-dimensional units, and initial longitude of periapsis and true anomaly both $\pi$ (for the interior MMRs considered in this study). This orbit will be symmetric about the $x$-axis and will also be periodic in the rotating Kepler problem (i.e., PCR3BP with $\mu = 0$). Thus, the method of perpendicular $x$-axis crossings can be used to numerically continue this Keplerian orbit to the true value of 
$\mu =  1.2150584270571545 \times10^{-2}$ for the Earth-Moon system; see, for example, Section 2.6.6.2 of Parker and Anderson \cite{parkerAnderson} for details of this method. The same method is then used to continue the resulting PCR3BP orbit through the rest of its orbit family, using the perpendicular orbit $x$-intercept as the continuation parameter. 

Once the periodic orbits in a family $k$:$k_m$ have been computed, the computation of their stable/unstable manifolds is carried out. In particular, we compute the intersection of these manifolds with the previously mentioned perigee Poincar\'e surface of section $\Sigma_C$. Such sections have been used by, e.g.,\ Ross and Scheeres \cite{RoSc2007} and Howell et al. \cite{howellDavisHaapala} as well; they have better transversality to the PCR3BP flow as compared to other commonly used sections such as $y = 0$. When using such a section, however, the periodic orbit intersection points with the section are not fixed points of the Poincar\'e map $P$, but become period-$k$ orbits under the map, as discussed in Section~\ref{sec:poincare}. This is because such an $k$:$k_m$ orbit passes through perigee $k$ times during one period, which takes approximately $k_m$ lunar sidereal periods. 

The portions of the periodic orbit stable/unstable manifolds lying in the chosen Poincar\'e section will correspond to 1D curves --- one curve for each of the period-$k$ points lying in the section. To help accurately compute the manifolds, we extended to the period-$k$ iteration orbit case the second author's previously developed parameterization method \cite{kumar2021journal,haroetal} for computing Taylor-series approximations of periodic orbit stable/unstable manifolds; this extension also incorporates many methods from the second author's previous work \cite{kumar2022} on computing manifolds of invariant tori. 
Although a full description is beyond the scope of this paper, in short, given an unstable resonant periodic orbit at Jacobi constant $C$, we solve for a function $W: \{ 0, \dots, k-1\} \times \mathbb{R} \rightarrow \mathcal{M}_C \subset \mathbb{R}^{4} $ such that, 

\begin{equation}  
    \label{invariancequationPCR3BP}   
    \phi_{\tau(i)}(W(i,s)) = W(i+1 \mod k, \lambda s) \quad i \in \{ 0, \dots, k-1\},   
\end{equation}
where $\phi_{t}(X)$ is the PCR3BP flow map of a point $X \in \mathcal{M}$ by time $t$, $\tau(i)$ is the time elapsed between the $i$th and $(i+1)$th periapsis passes of the periodic orbit being considered, and $\lambda$ is the $k$th root of the monodromy matrix eigenvalue corresponding to the stable/unstable manifold. Note that the $\phi_{\tau(i)}$ are \emph{not} Poincar\'e maps, but fixed-time maps. Equation \eqref{invariancequationPCR3BP} can be solved recursively by expressing $W$ as a set of Taylor series depending on the integer $i$,

\begin{equation}  
    \label{taylorSeries}   
    W(i, s) = \sum_{\bar{m}=0}^{\infty} W_{\bar{m}}(i) s^{\bar{m}} \quad i \in \{ 0, \dots, k-1\},
\end{equation}
\vspace{0.5mm}
\noindent where $W_{0}(i)=p_{i+1}$ are the periapsis period-$k$ points of the periodic orbit, and $W_{1}(i)$ are scaled eigenvectors of the periodic orbit monodromy matrix at each of its periapsis passages, with $W_{\bar{m}}(i)s^{\bar{m}}$ for $\bar{m}\ge2$ corresponding to higher-order terms in the stable/unstable manifold approximation. 

The $k$ curves parameterized by $W$ lie near but not on $\Sigma_C$, the periapsis section of interest. Thus, to finally compute the manifolds on the section, one simply numerically integrates dense grids of points from those curves either backwards or forwards to the section. Then, further applications of the Poincar\'e map $P$ either forwards or backwards in time are used to respectively globalize the full unstable and stable manifolds. As usual, for each fixed Jacobi constant value $C$, one can plot these Poincar\'e map manifolds of various orbits at that $C$ value using just 2D coordinates on $\Sigma_C$. Intersections of the 1D manifold curves, $\{W_{\pm}^{u,s}(p_i)\}$, will provide the geometry (e.g., PIPs, BIPs, homoclinic points, heteroclinic points) as discussed in Section~\ref{sec:poincare}.

\section{Semi-Analytical Approach to Identify Resonance Widths}
\label{sec:gallardo}

Standard tools have been developed in celestial mechanics to calculate the width (strength) and location of MMRs, under the perturbed-Hamiltonian formulation \cite{gallardo2006atlas,xWrM17,fNmM20}. They all invariably employ an expansion of the Hamiltonian around each resonant location and a {\it canonical transformation} to reduce the Hamiltonian to a system with only one single harmonic (i.e., normal-form reduction). While the mathematical developments here are straightforward, they can be algebraically quite complicated, especially for distant xGEO orbits of high eccentricity and inclination. 

The Hamiltonian describing the resonant dynamics near the  $k$:$k_m$ MMR is \cite{tG20},
\begin{align}
     \label{eq:gallardo}
    \mathcal{K} (a, \sigma) = -\frac{\mu_e}{2 a} - n_m \frac{k}{k_m} \sqrt{\mu_e a} - \mathcal{R} (a, \sigma), \quad \mathcal{R} (a, \sigma) = \frac{1}{2 \pi k_m} \int_0^{2 \pi k_m} R (\lambda_m, \lambda (\lambda_m, \sigma))\, \mathrm{d} \lambda_m,
\end{align}
where $\mu_e = G m_e$ is the Earth's gravitational parameter, $n_m = \sqrt{\mu_e/a_m^3}$ is the Moon's mean motion, $\lambda_m$ and $\lambda$ are the Moon and satellite mean longitudes respectively, and $R$ is the Moon's disturbing function,
\begin{align}
    R = \mu_m \left( \frac{1}{\lvert {\bm r}_m - {\bm r} \rvert} - \frac{{\bm r} \cdot {\bm r}_m}{r_m^3} \right).
\end{align}

The resonant disturbing function, $\mathcal{R} (a, \sigma)$, can be written as a series expansion of cosines whose critical arguments are of the type,
\begin{align}
    \sigma = k_m \lambda - k \lambda_m + \gamma,
\end{align}
where $\gamma$ is a slowly evolving angle defined by a linear combination of the longitudes of the ascending nodes $\Omega_{iner}$ and longitudes of perigee $\varpi_{iner} = \Omega_{iner} + \omega_{iner}$ of the satellite and Moon.

Gallardo \cite{tG19,tG20} uses a numerical computation for the averaging in Eq.~\eqref{eq:gallardo}, assuming that the Moon's orbit is Keplerian and lies in the ecliptic plane, and taking for the spacecraft the semi-major axis corresponding to the nominal position of the resonance, and assuming the satellite's eccentricity, inclination, periapsis, and node are fixed during the period of time in which the integral is calculated. 
This is justified in the asteroid case by the otherwise slow evolution timescale of $(e, i, \omega, \Omega)$, as compared to the oscillations of $a$ and $\sigma$. 

\begin{figure}[hbt!]
\centering
\includegraphics[width=0.9\linewidth]{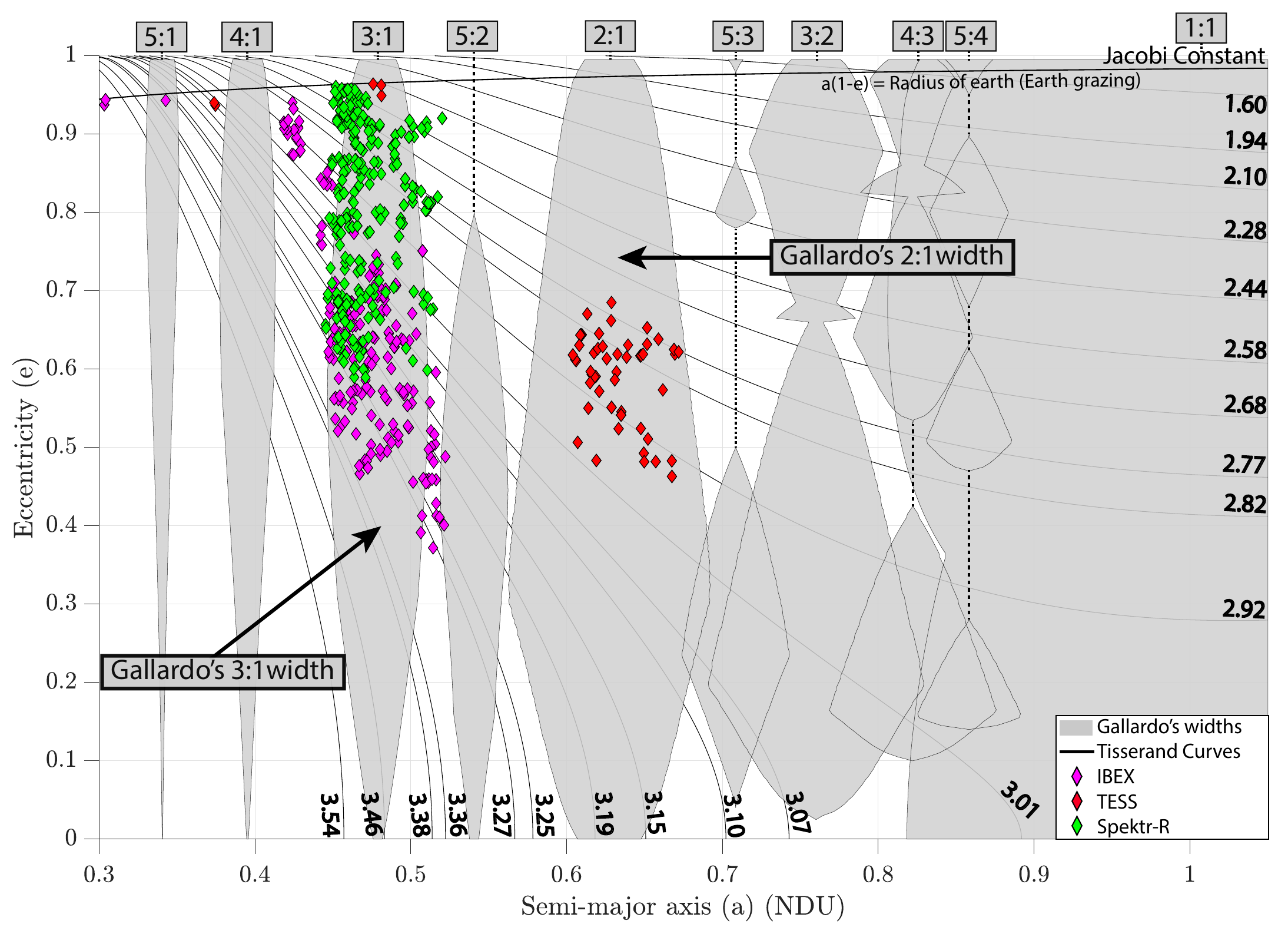}
\caption{\label{fig:TLEs_Gallardo}TLEs of cataloged xGEO space objects, TESS, IBEX and Spektr-R, projected onto the semi-major axis-eccentricity $(a,e)$ plane, superimposed on Gallardo's analytical approximation of widths computed assuming coplanar Moon's orbit.}
\end{figure}

For the Earth-Moon system, the resonance widths calculated using Gallardo's algorithm are shown in
osculating semi-major axis-eccentricity $(a,e)$ space in Fig.~\ref{fig:TLEs_Gallardo}, using the mean orbit of the Moon (i.e., $a_m = 383397.7725$ km, $e_m = 0.055545526$) as input parameters for the perturber \cite{jS94}. The satellite's inclination was set to zero and the $R_H$ (the Moon's Hill radius) factor was set to 2.0. The phase-space structure of the widths of 3:1 and 2:1 resonances generally resemble those in the small-body context (e.g., hour-glass shapes) \cite{lLrM19}, but the 1:1 (co-orbital resonance) is dramatically over predicted with its width encompassing the entire lunar Hill sphere. The other resonances appearing in Gallardo's ``atlas'', limited herein to order 5, will not be discussed further. The time histories of notable resonant satellites (IBEX, Spektr-R, and TESS), obtained from TLE data (www.space-track.org, Assessed 10 Mar. 2023), are overlaid; it can be observed that TESS lies within the predicted libration zone of the 2:1, but the predicted 3:1 width does not fully encompass the other spacecraft.

\section{Direct Identification of Resonance Widths and Chaotic Zones}
\label{sec:zones}
 
\begin{figure}[hbt!]
\centering
\includegraphics[width=1\linewidth]{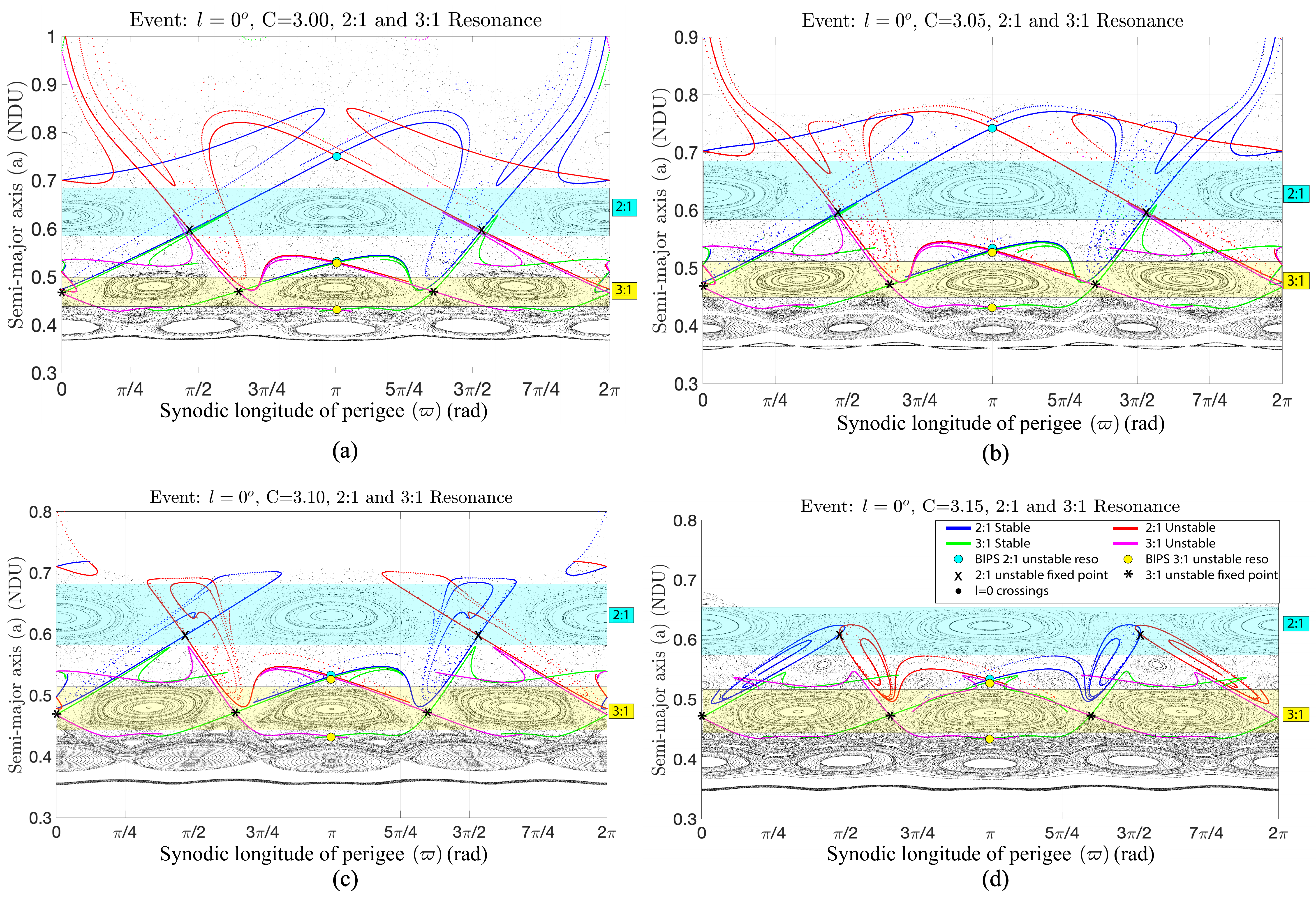}
\caption{\label{fig:2:1 & 3:1 resonance} Poincar\'e sections $\Sigma_C$ for (a) $C=3.00$, (b) $3.05$, (c) $3.10$, and (d) $3.15$, depicting stable resonance widths and chaotic resonance zones of the stable 2:1 and 3:1 MMRs.}
\end{figure}

We calculate and plot intersections of numerous PCR3BP orbits with the Poincar\'e section $\Sigma_C$ for a range of Jacobi constants, revealing various resonances; see the $(\varpi,a)$-plane in
Fig. \ref{fig:2:1 & 3:1 resonance}, where, recall, $\varpi$ is the longitude of perigee in the rotating frame.
The most prominent resonances are the 2:1, 3:1, and 4:1 MMRs ( 4:1 MMR is identified but will not be discussed further.). The stable resonances are particularly apparent as ``islands'' of concentric closed curves surrounding center points that are stable period-2, period-3, and period-4 orbits, respectively. The closed curves surrounding the stable fixed points
are stable quasi-periodic librational tori (recall Section~\ref{sec:resonances}). 
Surrounding these resonance islands are regions of chaos. 
These regions are not featureless. 
Instead, the template of the motion is given by the corresponding  unstable MMR periodic orbits, which appear as saddle-type period-$k$ points on the Poincar\'e section $\Sigma_C$.
Using the method of Section~\ref{sec:manifolds}, these periodic orbits and their stable and unstable manifolds are computed.  The stable and unstable manifolds intersect to form homoclinic and heteroclinic tangles, which provide the paradigm with which to understand chaos in the CR3BP \cite{Poincare1892,KoLoMaRo2000}.

The region of existence of a particular mean-motion resonance can be identified by examining the Poincar\'e section, which also reveals the strength (width) of the resonance. Resonance widths were defined in Section~\ref{sec:poincare} (see Fig.~\ref{fig:resonance_zone_stable_and_bips}), and are shown for the 3:1 and 2:1 MMRs on Poincar\'e maps computed for $C=3.00,3.05,3.10,3.15$ in Fig.~\ref{fig:2:1 & 3:1 resonance}.

The influence of each resonance region extends beyond the outermost stable resonant librational torus regions to the ``separatrix'', the boundary formed by the intersection of stable and unstable manifolds described in Section~\ref{sec:poincare} (see Fig. \ref{fig:two-resonance-regions}). Focusing on the MMR bands, we use the methods of Section~\ref{sec:poincare} to 
identify BIPs and designate the boundary of the (chaotic) resonance region. BIPs defining the chaotic zones are depicted on the same Poincar\'e maps in Fig. \ref{fig:2:1 & 3:1 resonance}. For $C\geq 3.10$ the upper 2:1  BIP disappears. In fact, we see a ``swirling'' of manifolds that is a consequence of interaction with the L1 Lyapunov stable manifold cut resulting in trajectories ``exiting'' the
Poincaré section in the Earth realm, entering the Moon realm, and then re-emerging in the Earth realm. This phenomenon will be explored in future work.

\begin{figure}[hbt!]
\centering
\includegraphics[width=0.9\linewidth]{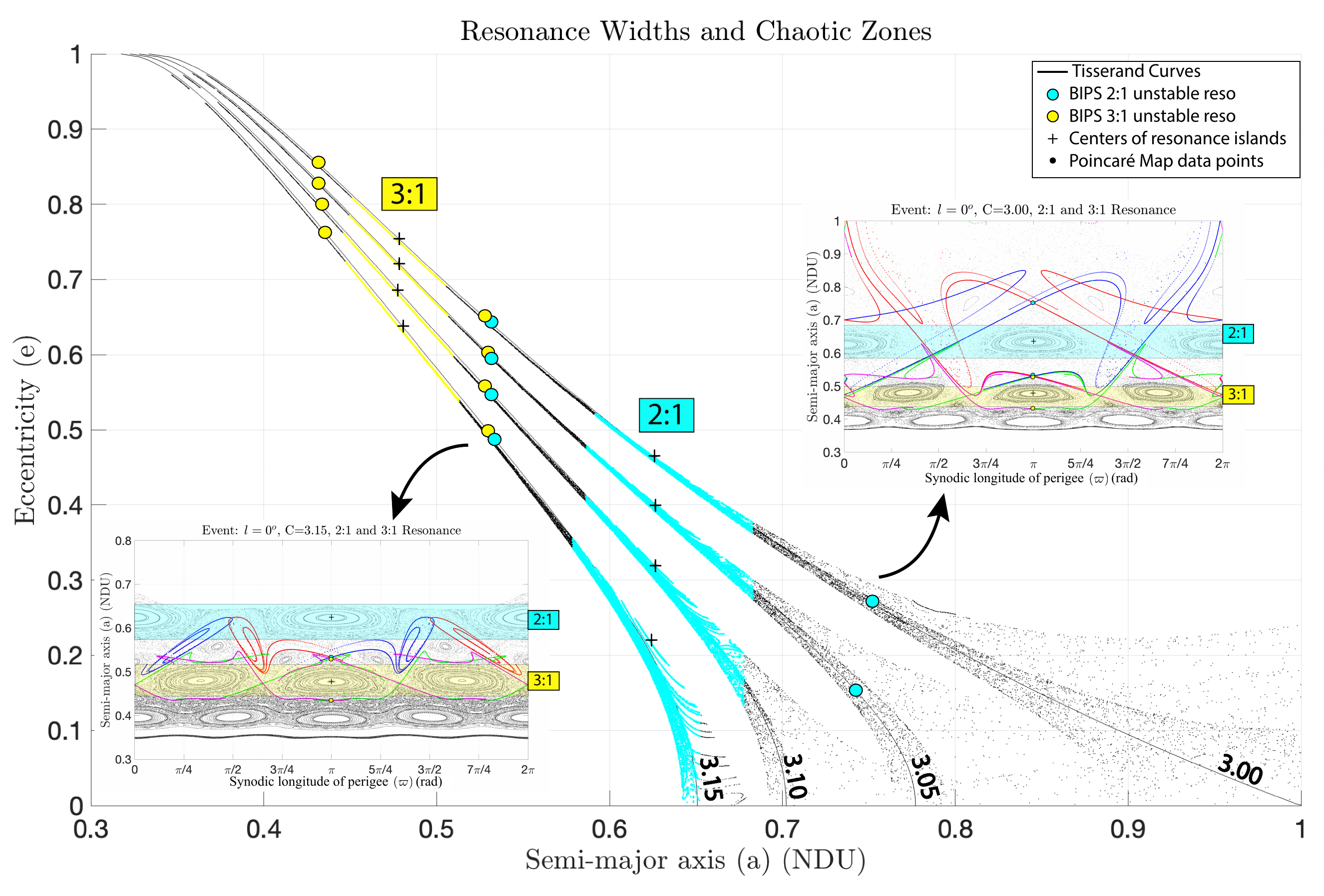}
\caption{\label{fig:width4}Resonance widths and chaotic resonance zones (delineated by BIPs) for 2:1 and 3:1 resonance for $C=3.00, 3.05, 3.10, 3.15$ shown in the $(a,e)$ plane.}
\end{figure}

While the resonances are identified and visualized in the $(\varpi,a)$ space, we also aim to represent them in the $(a, e)$ plane for direct comparison with space-object distributions and the semi-analytical resonance widths shown in Fig.~\ref{fig:TLEs_Gallardo}. 
The points from the Poincar\'e section, which exist in the 3-dimensional $(a, e, g)$ space for each $C$, can be projected onto the $(a, e)$ plane, as illustrated in Fig.~\ref{fig:width4}. 
In this figure, the stable resonance widths and the BIPs that bound the resonance zones are clearly delineated. 
As discussed in Section~\ref{sec:CR3BP}, Tisserand curves provide an approximation of the projection of the energy surface $\mathcal{M}_C$ onto the $(a, e)$ plane.
For each Jacobi constant $C$ in Fig.~\ref{fig:width4}, we include the corresponding Tisserand curve to highlight the level of agreement, particularly for smaller semi-major axes.

\begin{figure}[hbt!]
\centering
    \begin{tabular}{cc}
	\hspace{-0.7cm}\includegraphics[width=0.49\linewidth]{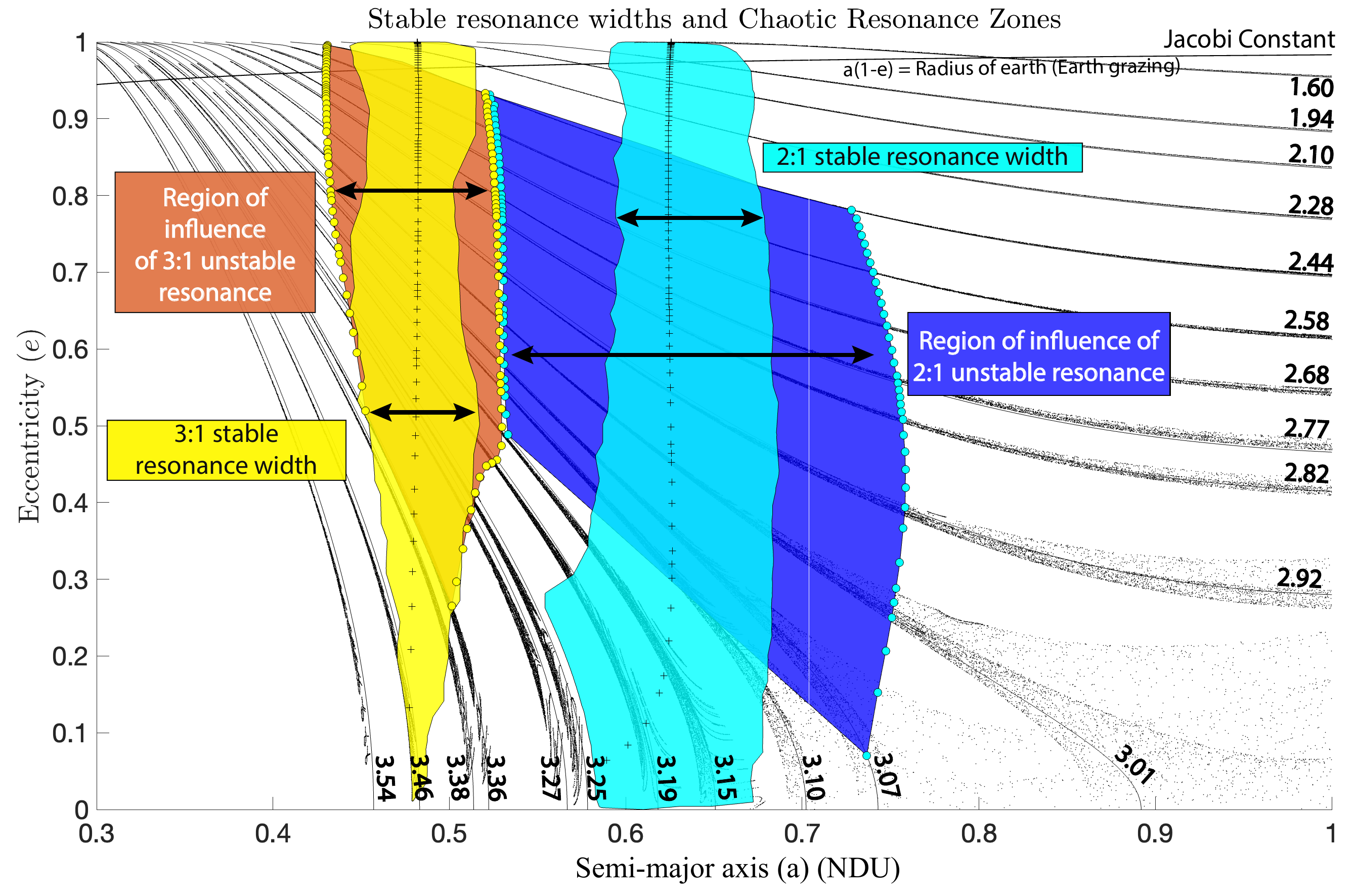} &
   \hspace{-0.3cm} \includegraphics[width=0.53\linewidth]{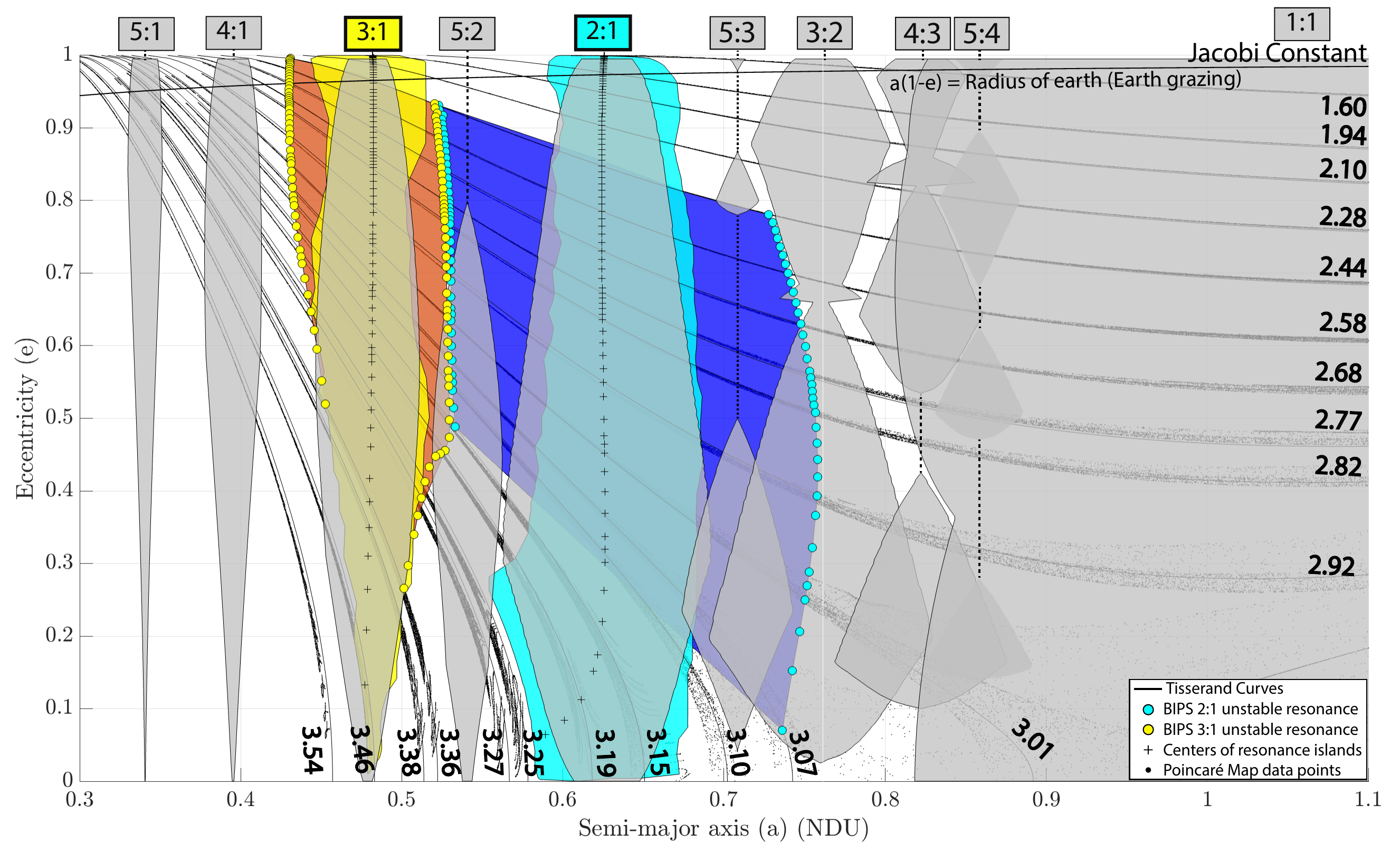}\\ 
        (a)  & (b) 
     \end{tabular}

\caption{(a) Resonance widths and chaotic resonance zones for 2:1 and 3:1 resonance for $C=1.6$ to $C=3.54$ with increments of $\Delta C = 0.02$. (b) PCR3BP widths overlapped on Gallardo's widths.}
\label{fig:width_overlap}
\end{figure}

The results in Fig. \ref{fig:width4} illustrate our method, but the number of Jacobi constants is sparse.
To get a fuller picture of the (stable) resonance widths and the (chaotic) resonance regions, 
Poincar\'e sections were computed for Jacobi constant values $C$ ranging from $1.60$ to $3.54$ in increments of 
$\Delta C = 0.02$. As shown in Fig. \ref{fig:width_overlap} (a), the Poincar\'e section points were projected onto the $(a,e)$ plane, marking resonance widths and chaotic resonance zones (via BIPs) for the 2:1 and 3:1 resonances. The family of prograde stable 3:1 resonant periodic orbits exists for $C=2.10$ to 3.47, with trajectories resulting in Earth collisions for $C\leq2.44$. Similarly, the prograde stable 2:1 resonant periodic orbit family begins at $C=1.6$, with Earth-impacting orbits occurring for $C\leq1.94$. As our current manifold computation methods do not incorporate Levi-Civita regularization, accurate manifold computations for unstable resonant orbits approaching or colliding with Earth were not possible. Consequently, manifolds are plotted starting only from $C=2.44$.

The stable widths obtained by using the full PCR3BP model are comparable to those computed using Gallardo’s algorithm, as seen in Fig. \ref{fig:width_overlap} (b). However, Gallardo’s widths are computed from the resonance center (stable equilibrium) to the maximum libration, capturing only the stable libration zones. This method neglects the deformation of resonance separatrices due to interactions with neighboring resonant harmonics, a significant effect at higher mass ratios. Consequently, when compared to the chaotic resonance zones indicated by the BIPs, Gallardo’s algorithm substantially underestimates the true region of influence of 2:1 and 3:1 unstable resonances.

Specifically, the PCR3BP stable width for the 2:1 resonance appears larger than Gallardo’s approximation for near-zero and near-unity eccentricities, but are slightly smaller for eccentricities in between. For eccentricities approaching unity, higher-order resonances --- also known as ``sticky islands'' --- form around the primary resonance island. The invariant curves start surrounding these sticky islands and contribute to the increased widths at high eccentricities, a phenomenon Gallardo’s algorithm does not capture. Moreover, the shape of the 2:1 resonance does not show the characteristic shape seen in other CR3BP-based resonance-width computations \cite{xWrM17,malhotra2020divergence} --- a largest width near some eccentricity $0.3 \le e\le 0.7$ tapering to significantly smaller widths as $e$ approaches 0 and 1.
We note that earlier work tended to assume that resonance widths should be along lines of constant semi-major axis in the $(a,e)$ plane \cite{xWrM17}, which does not consider the `tilt' of the energy surface (as approximated by Tisserand curves).

The region of influence of 2:1 and 3:1 unstable resonance regions, as represented by BIPs, nearly touch in the region between them, indicating possible heteroclinic connections. The stable resonance zones of 2:1 and 3:1 span approximately 28,328 km and 20,281 km in the semi-major axis, respectively, when averaged across all Jacobi constants. The regions of influence of 2:1 and 3:1 unstable resonances respectively span approximately 83,982 km and 34,619 km in semi-major axis, when averaged across all Jacobi constants. These regions encompass numerous higher-order resonances, suggesting the potential for multiple free transfers between different order resonances.

\begin{figure}[hbt!]
\centering
\includegraphics[width=1\linewidth]{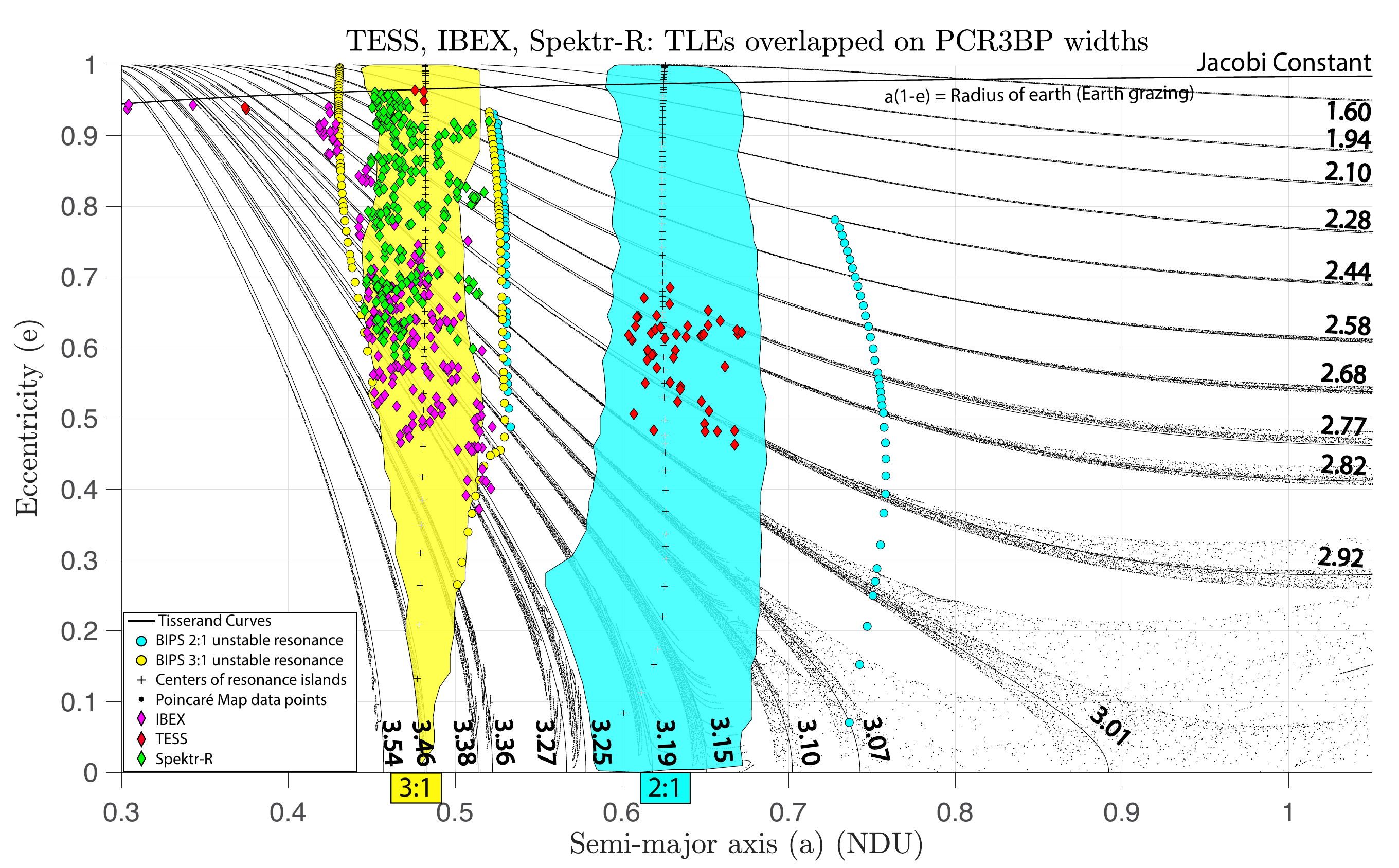}
\caption{\label{fig:tle_overlapped} TLEs of xGEO space objects (from Fig. \ref{fig:TLEs_Gallardo}) superimposed on the PCR3BP-based widths. }
\label{fig:TLEs_onwidths}
\end{figure}

As done previously in Fig. \ref{fig:TLEs_Gallardo}, time histories of IBEX, TESS and Spektr-R are overlaid in  Fig.~\ref{fig:tle_overlapped}.
The PCR3BP resonance widths effectively encompass several spacecraft, such as IBEX, TESS, and Spektr-R. In contrast, the narrow semi-analytical resonance zone approximation inadequately represents the spacecrafts' locations. Notably, the PCR3BP resonance zones in $(a,e)$ can accurately discern the resonant dynamical character of even non-coplanar spacecraft like IBEX, underscoring its utility in providing fundamental insights into spacecraft dynamics in cislunar space. Utilizing these resonance widths facilitates the determination of whether a spacecraft is within a stable or chaotic orbital regime, thereby aiding in mission analysis and prediction of its future orbital evolution.

For instance, IBEX's nominal orbit exhibited chaotic behavior due to lunar perturbations, prompting a transfer to a stable 3:1 resonant orbit with the Moon \cite{Dichmann2013}. A comprehensive understanding of chaotic dynamics in such environments would have been valuable during the mission's early design phases. This understanding has partly informed the orbital design of missions like TESS \cite{Ricker2014}. Although significant challenges remain in comprehending lunar secular and mean-motion resonances, progress in discerning the influence of specific resonances, such as the 2:1 and 3:1 MMRs, through the PCR3BP, contributes to advancing our understanding of these important lunar resonances.

\section{Comparison with the Ephemeris Model}
\label{sec:ephem}

\begin{figure}[hbt!]
\centering
    \begin{tabular}{ccc}
	\includegraphics[width=0.32\linewidth]{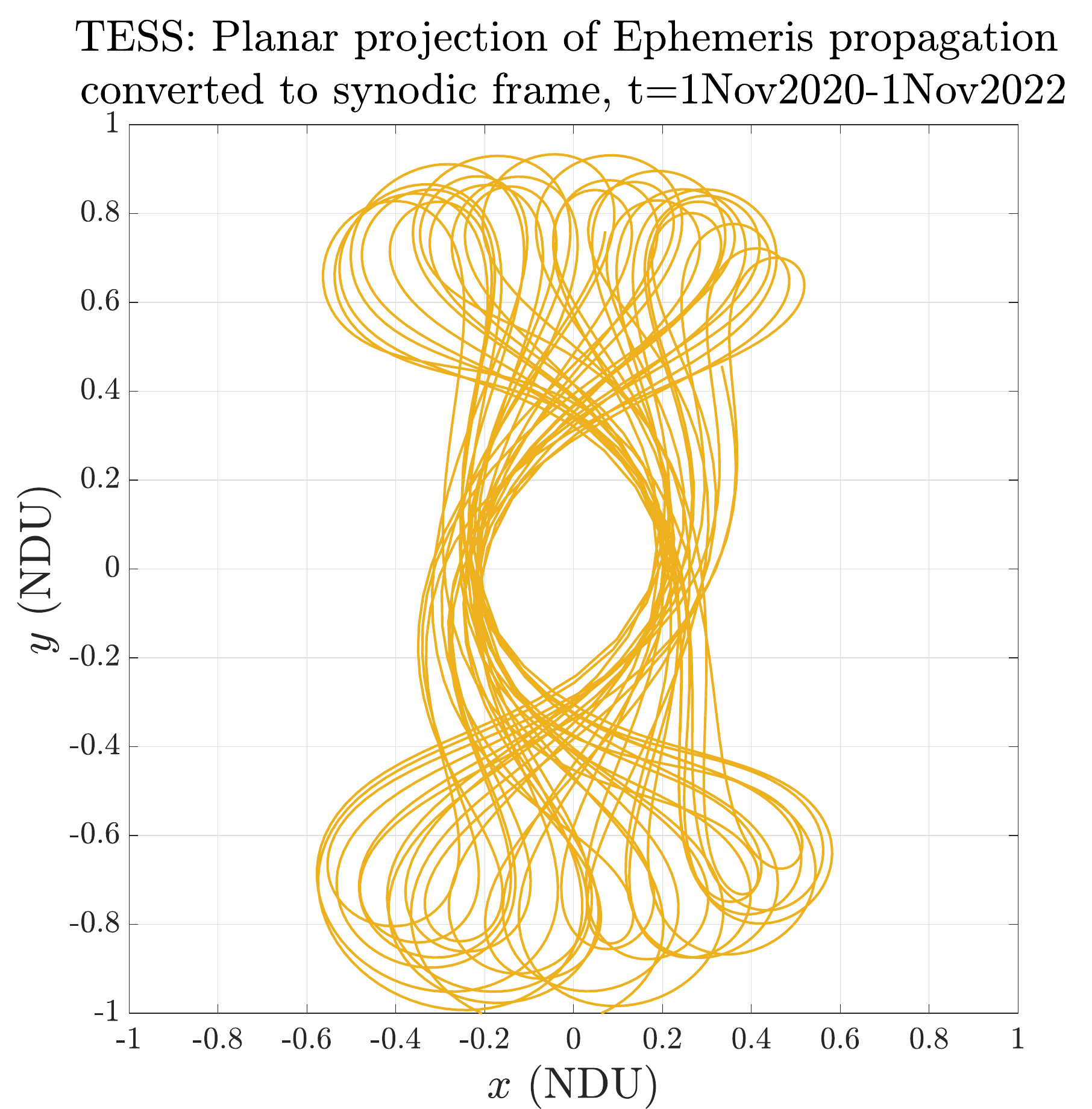} &
    \includegraphics[width=0.32\linewidth]{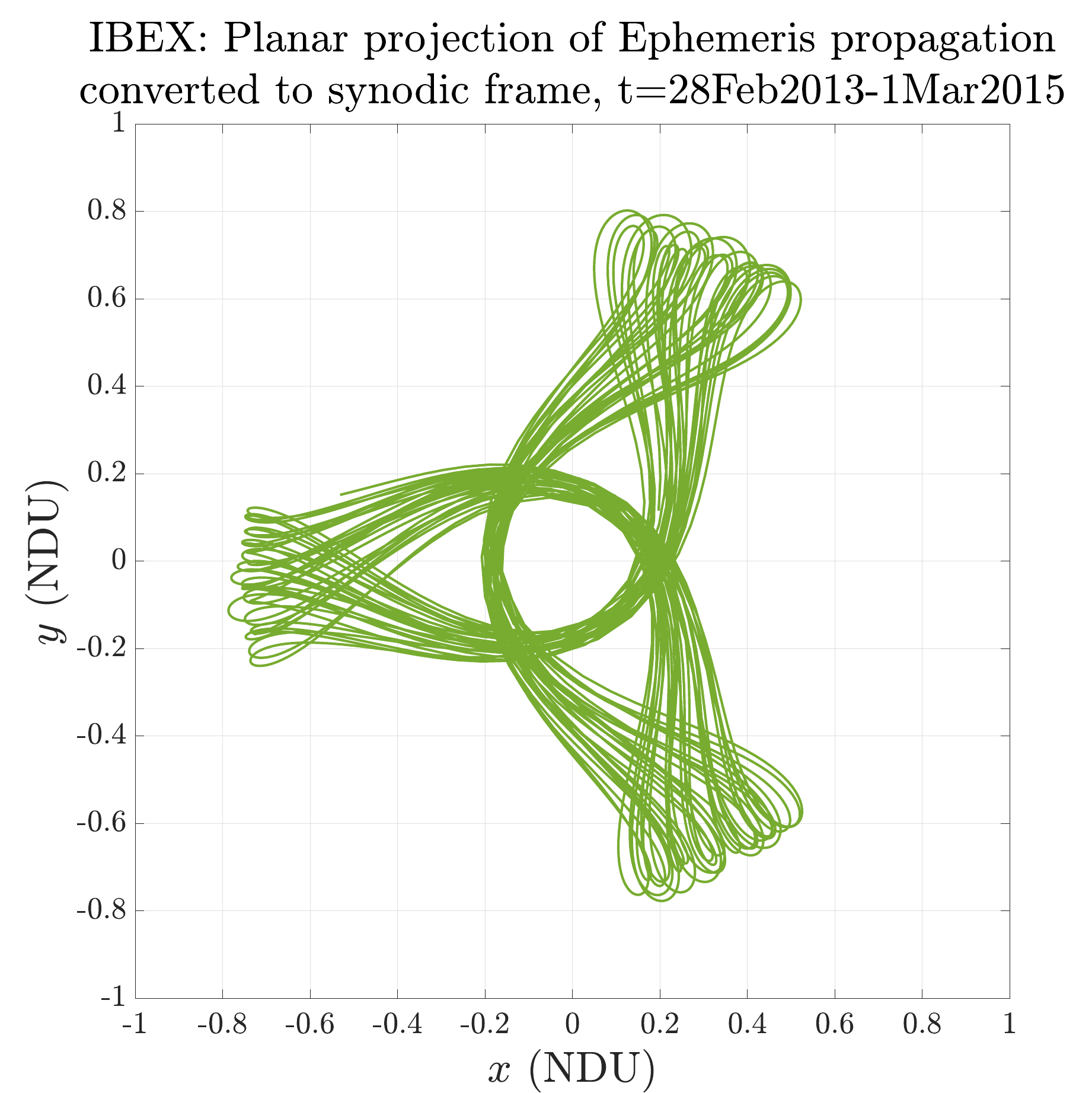} &
    \includegraphics[width=0.32\linewidth]{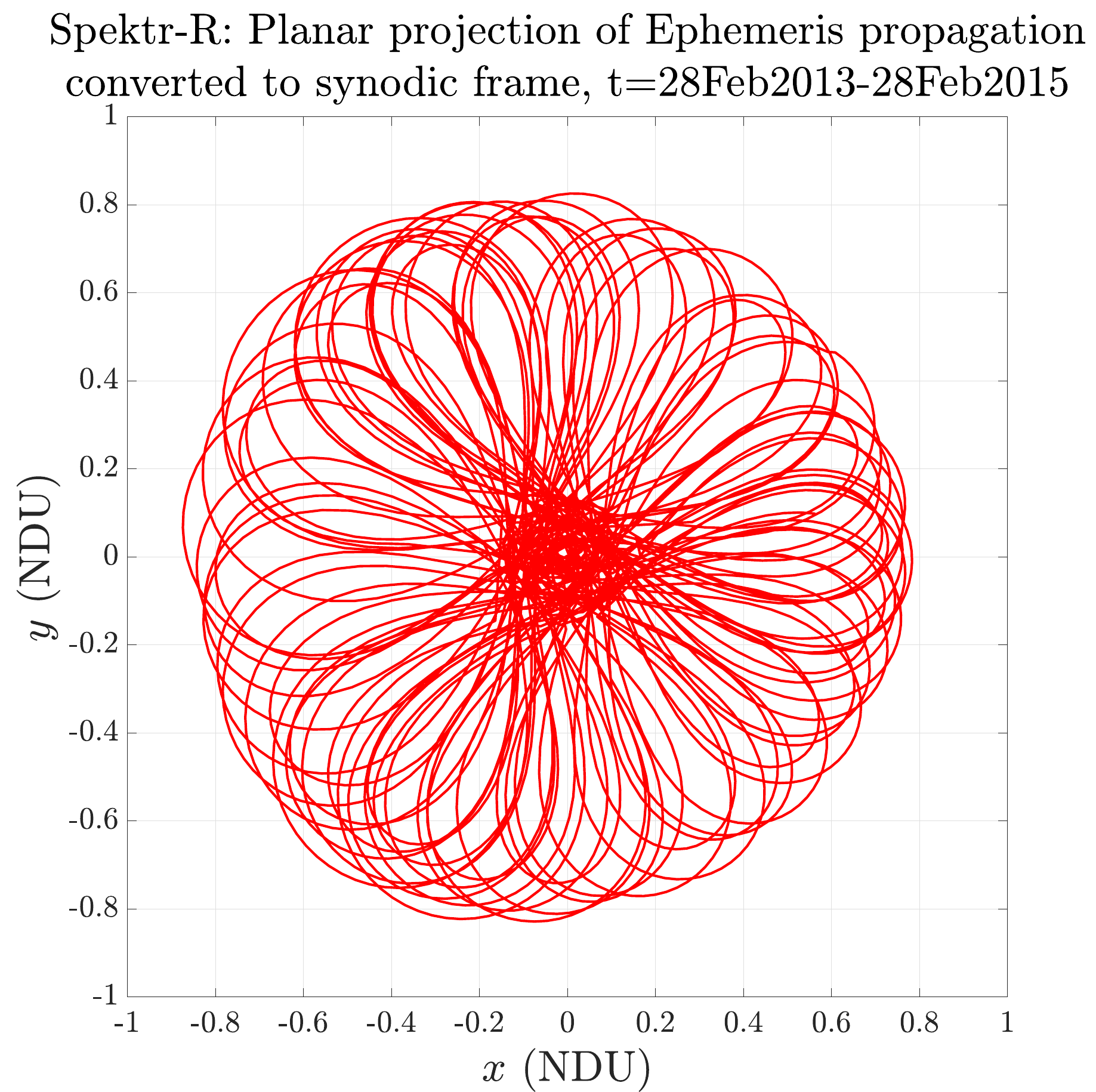}\\ 
        (a)  & (b) & (c)
     \end{tabular}

\caption{\label{fig:traj_individual} Trajectories propagated using JPL Horizons (TESS, Spektr-R) and Cowell's 4-body ephemeris model (IBEX), converted into synodic states and projected onto the Earth-Moon $xy$ plane.}
\end{figure}

Figure~\ref{fig:traj_individual} shows planar projections of ephemeris propagations of resonant spacecraft: TESS is situated in a stable 2:1 MMR, mostly occupying a larger quasi-periodic torus; IBEX is located in a stable 3:1 MMR; and Spektr-R navigates in unstable regimes, thus having a chaotic trajectory.

 The orbital trajectories of TESS, IBEX, and Spektr-R, as described by their TLEs, predominantly fall within the stable resonance widths of the PCR3BP, as illustrated in Fig.~\ref{fig:TLEs_onwidths}. TLEs provide a snapshot of a satellite's orbital elements at specific epochs, typically at the ascending node. Therefore, their comparison with PCR3BP stable resonance widths can serve as a preliminary validation. However, TLEs are inherently sporadic and do not accurately represent osculating orbital elements at or near the perigee. To address this limitation, the trajectories of TESS, IBEX, and Spektr-R are propagated using high-fidelity Sun-Earth-Moon-satellite ephemeris models --- Horizons and Cowell's 4-body propagator ---enabling the determination of perigee crossings to map onto the PCR3BP widths  $(a,e)$  plane and on the Poincar\'e maps'  $(\varpi,a)$ plane.

In the spatial CR3BP, the synodic longitude of perigee is defined as the sum of the inertial argument of perigee added to the synodic angle between the Earth-Moon line and the ascending node direction ($\varpi = \Omega + \omega$).

For TESS and Spektr-R, trajectory propagation is conducted using JPL Horizons data for 2-year intervals, from initial epoch 1 November 2020 for TESS and 28 February 2013 for Spektr-R. A data extraction time step of 10 minutes is selected after confirming negligible variations in near-perigee states with finer data extraction time steps up to 1 minute. Data points are extracted with respect to the International Celestial Reference Frame (ICRF), centered at the Earth and ecliptic reference plane. Near-perigee crossings are identified based on minimum distance and the condition $\mathbf{r} \cdot \mathbf{v} = 0$. The inertial perigee states are then converted to synodic states using the methodology described in Appendix B. The Jacobi constant (for each state), calculated using spatial states, is averaged over all states to account for its variation in the high-fidelity model, after which a 2-dimensional plot of Poincar\'e map orbit points is computed at that same spatial Jacobi constant value. The spacecraft synodic perigee crossings converted into geocentric osculating orbital elements are then mapped onto the Poincar\'e map plot and PCR3BP widths. Additionally, spatial CR3BP propagation of the initial Horizons/Cowell's perigee state is performed and subsequent perigee crossings are mapped onto the Poincar\'e map for comparison. For IBEX, Horizons data is unavailable, necessitating trajectory propagation using Cowell's 4-body propagator over 2 years (28 February 2013 to 1 March 2015), with the initial condition for IBEX being that reported by \cite{Dichmann2013}. Perigee states are identified using MATLAB's event functionality, converted to synodic states, and mapped similarly onto the Poincar\'e map and PCR3BP widths.

\begin{figure}[hbt!]
\centering
    \begin{tabular}{ccc}
	\includegraphics[width=0.30\linewidth]{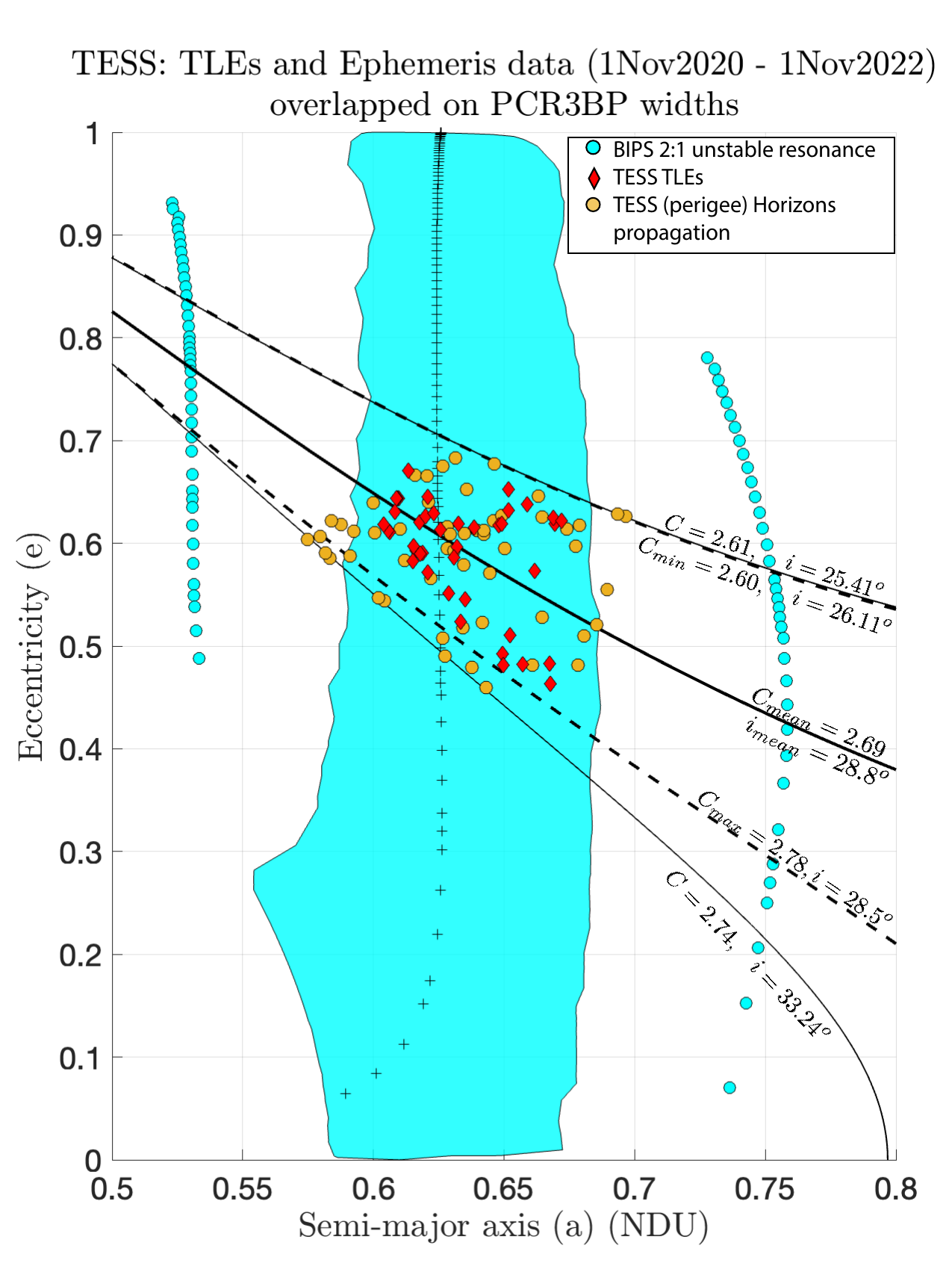} &
    \includegraphics[width=0.30\linewidth]{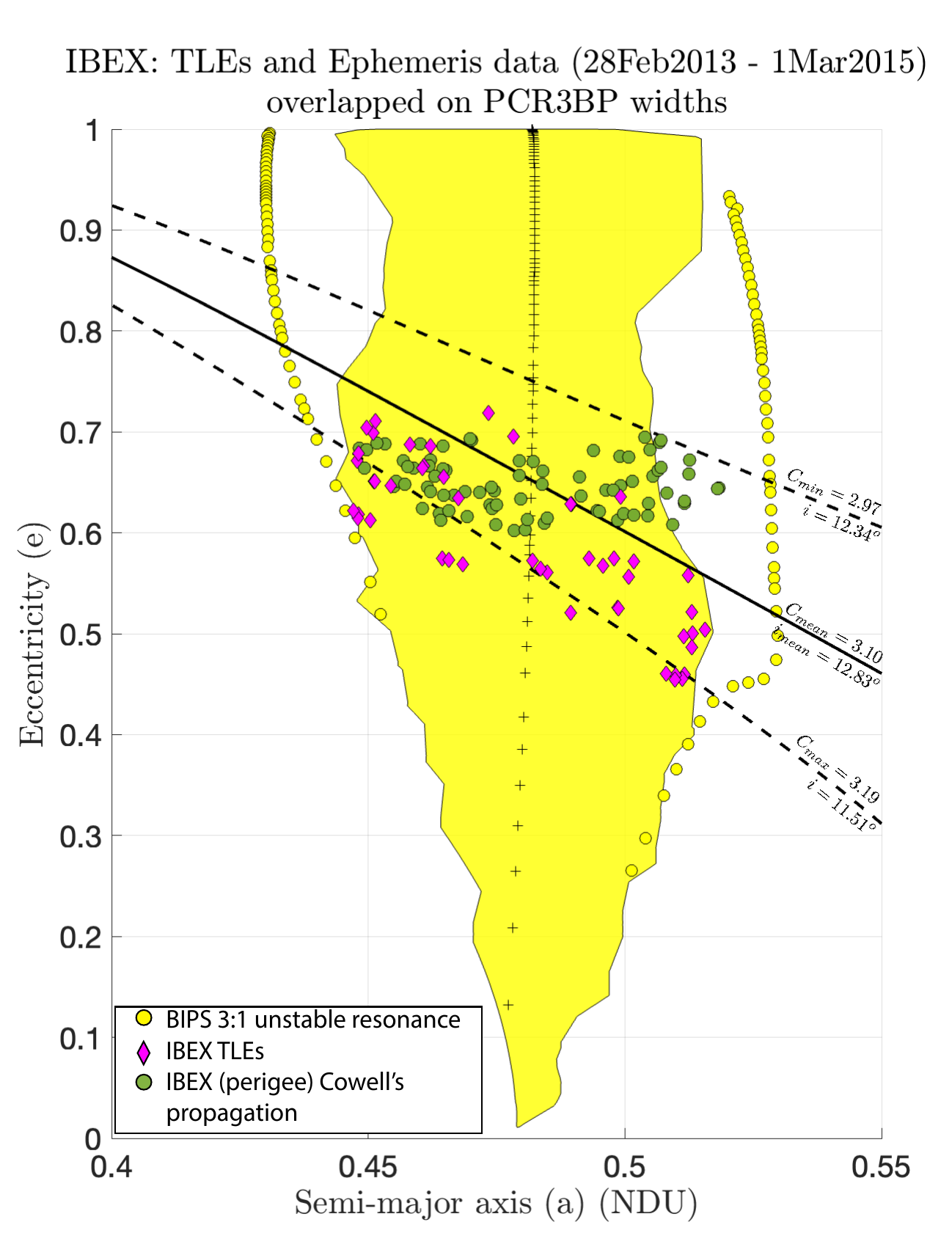} &
    \includegraphics[width=0.32\linewidth]{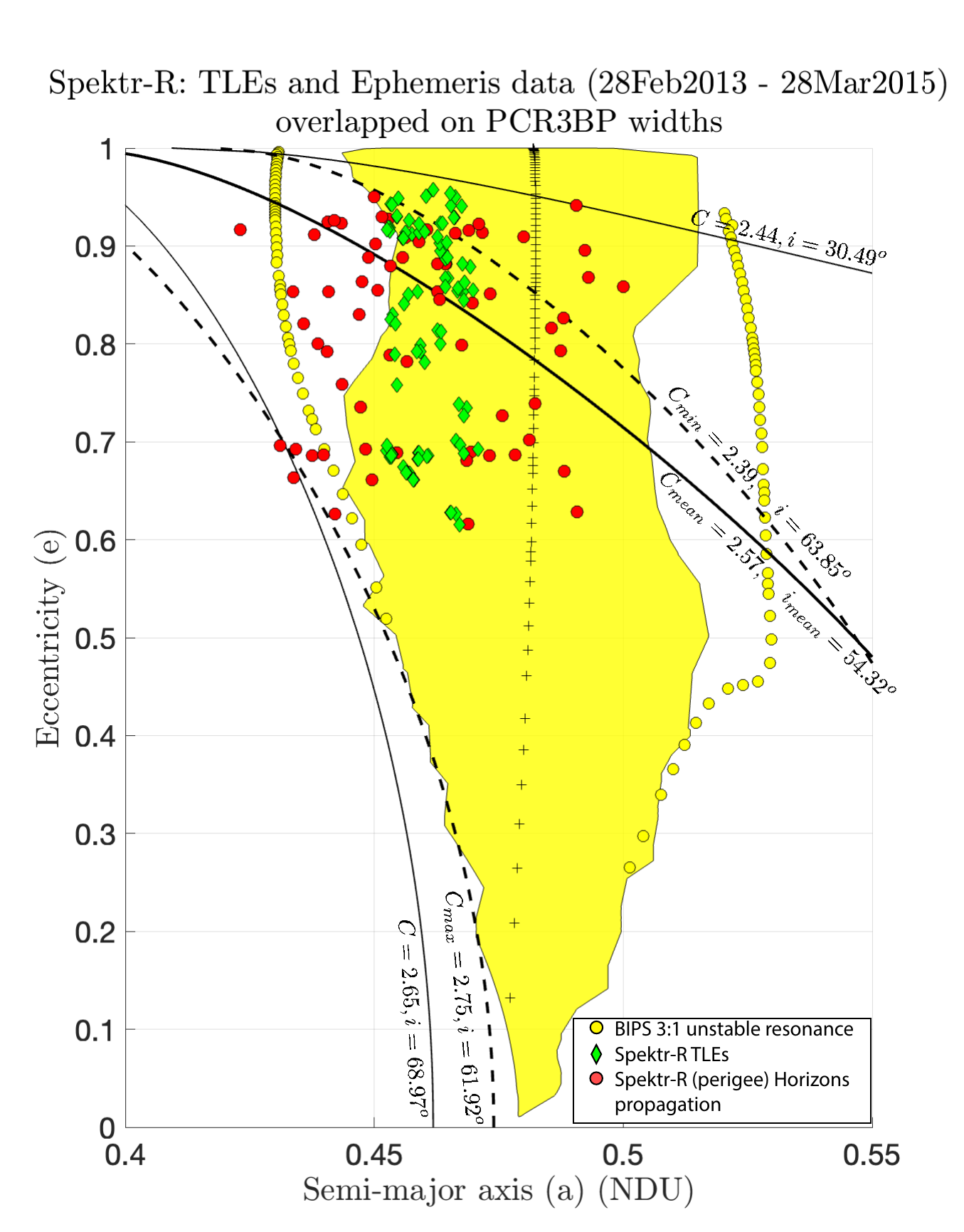}\\ 
        (a)  & (b) & (c)
     \end{tabular}

\caption{\label{fig:widths_individual} (a) TESS, (b) IBEX, (c) Spektr-R: TLE and ephemeris perigees overlapped on PCR3BP widths, with inclined Tisserand curves for max/min (dashed) and mean (bold solid) Jacobi constants and to capture spread (light solid).}
\end{figure} 

Fig.~\ref{fig:widths_individual} compares TLEs and the perigee states derived from high-fidelity propagation of TESS, IBEX, and Spektr-R over two years. The results indicate that the perigee states of TESS and IBEX, which are associated with stable 2:1 and 3:1 resonances, respectively, largely align with the stable PCR3BP resonance widths. However, some data points fall outside the widths attributable to including higher-fidelity effects in the propagation. Despite these deviations, the PCR3BP widths approximate the underlying fundamental dynamical framework. Spektr-R, on the other hand, exhibits perigee states predominantly within the chaotic resonance zones of the unstable 3:1 resonance. This is consistent with its known dynamical behavior caused by significantly high ecliptic inclination, which permits von Zeipel-Lidov-Kozai oscillations \cite{Amato2020,Shevchenko2017} and other secular resonant phenomena, thus navigating unstable regimes.

The Tisserand curves, incorporating the effect of inclination with respect to the Earth-Moon plane, have also been plotted in Fig.~\ref{fig:widths_individual} for the maximum and minimum Jacobi constants computed for the two-year perigee data points. Additionally, a mean Jacobi constant Tisserand curve, combined with the mean inclination is presented. This reveals that most data points for TESS and IBEX predominantly reside within the established Tisserand boundaries for extreme Jacobi constants. Occasional deviations observed can primarily be attributed to the inherent limitations of Tisserand curves in approximating the true energy surface. Furthermore, since the inclination of individual trajectory points varies significantly, instances are observed where an intermediate Jacobi constant Tisserand curve corresponding to a specific data point energy, but coupled with a higher inclination, may subsequently be plotted outside the extreme Jacobi constant Tisserand boundaries, as is particularly noticeable in the TESS case for ($C=2.74, i=33.24^o$).

\begin{figure}[hbt!]
\centering
    \begin{tabular}{cm{18cm}cm{18cm}cm{18cm}}
	(a) & \includegraphics[width=0.75\linewidth]{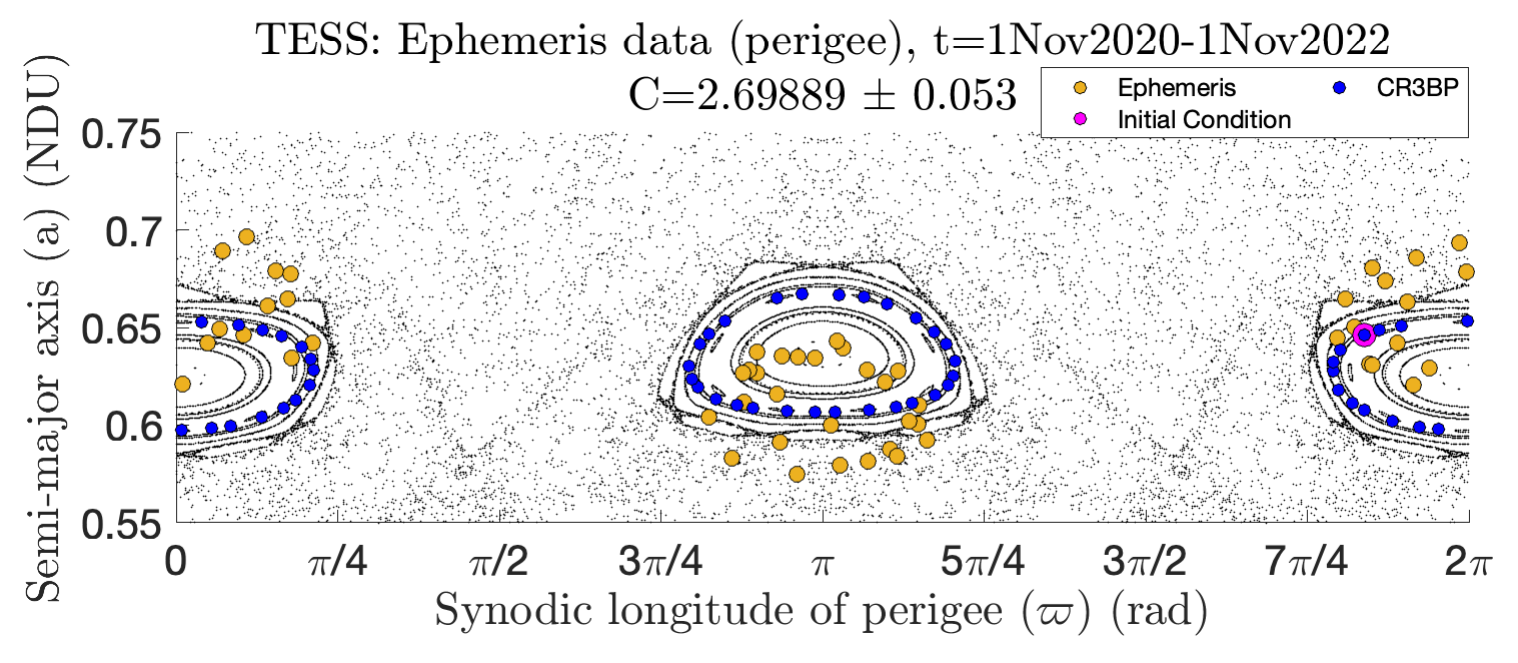}  \vspace{-5mm}\\
    (b) & \includegraphics[width=0.75\linewidth]{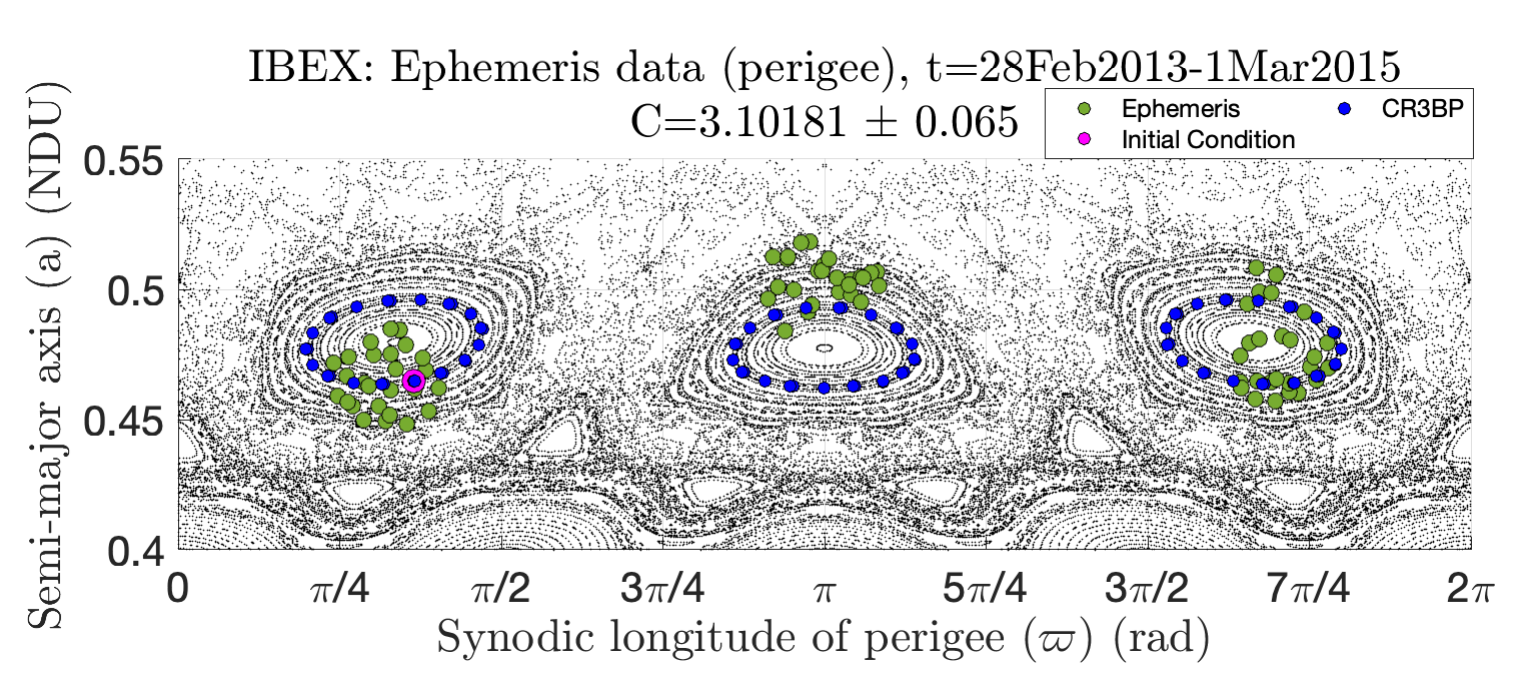} \vspace{-5mm}\\
    (c) & \includegraphics[width=0.75\linewidth]{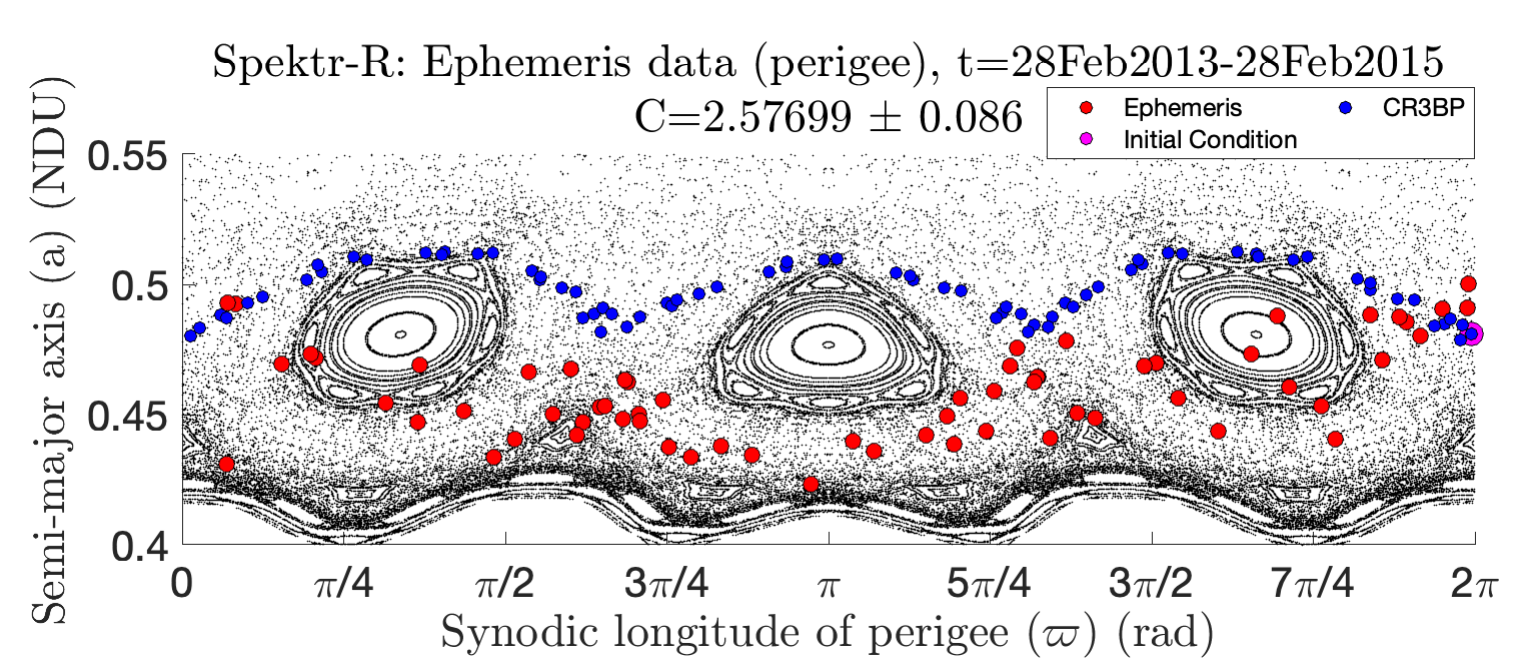}\\
    \end{tabular}

\caption{\label{fig:map_individual} (a) TESS, (b) IBEX, (c) Spektr-R: CR3BP propagation and Horizons data (perigee) overlaid on Poincar\'e map, C = 2.69889, C=3.10281, C=2.57688 respectively.}
\end{figure} 

For IBEX, it is noted that the TLEs during the specified observation period are quite sporadic, contributing to their deviation from the predicted ranges of Tisserand curves. Spektr-R presents a distinctive case characterized with substantial variations in orbital inclinations. Notably, the perigee point corresponding to $i=30.49^o$ has an $e>0.9$, whereas data points corresponding to $i=68.97^o$ have $e<0.7$. This observed trade off between eccentricity and inclination illustrates the effect of secular resonances.

Mapping perigee crossings to the $(a,e)$ plane alone is insufficient to confirm stability of a satellite, as observed in the case of Spektr-R. Instead, the $(\varpi,a)$ plane is also examined to verify alignment of perigee crossings with stable resonant islands on the Poincar\'e map. Thus, high-fidelity propagated perigee states of these three satellites are mapped onto the respective Poincar\'e maps computed for averaged Jacobi constants, as shown in Fig.~\ref{fig:map_individual} along with the spatial CR3BP propagation of their initial perigee state (marked). 

The perigee mappings obtained from spatial CR3BP propagations of initial states for TESS and IBEX (denoted in blue) align closely with libration regions, or resonant islands, in the planar Poincar\'e map represented in the $(\varpi,a)$ plane. This correlation suggests that planar dynamics can effectively inform spatial dynamics when analyzed within the $(\varpi,a)$ parameter space, a relationship worthy of further study, and under investigation. Additionally, sequential perigee mappings corresponding to the two resonant islands exhibit an approximately conserved area on the $(\varpi,a)$ plane. This behavior reflects the connection between orbital elements and their corresponding Delaunay variables, thus adhering to the integral invariants of Poincar\'e-Cartan (see Appendix C). The mappings from the CR3BP for TESS and IBEX conform closely to expected stability profiles, indicating that TESS is situated near the separatrix of the 2:1 resonance, bordering regions of chaotic dynamics, whereas IBEX occupies a stable inner torus within the 3:1 resonance. The high-fidelity propagated perigee states of TESS and IBEX cluster around the stable islands of the 2:1 and 3:1 resonances, respectively. Discrepancies in shape of island clusters between high-fidelity states and the PCR3BP stable islands are expected due to the inclusion of the effect of the eccentricity of the Moon and the effect of the Sun. 

Conversely, the CR3BP propagation of Spektr-R exhibits a trajectory in the unstable region of influence of the 3:1 resonance, i.e., within the chaotic zone. The perigee states from ephemeris data are dispersed in a similar manner, corroborating its navigation within an unstable dynamical regime during the studied period. This analysis reinforces the utility of the PCR3BP framework to delineate satellite dynamics and resonance stability, which is validated by high-fidelity propagation analysis, highlighting the Earth-Moon system as an astrodynamics laboratory \cite{alessi2019earth}.

\section{Conclusion}
\label{sec:conclusion}

Poincar\'e sections at perigee provide a detailed depiction of the dynamical structure within cislunar space, highlighting both stable islands and chaotic regions, all depicted in geocentric orbital elements which remain immensely useful for cislunar space. Our results highlight the pronounced impact of the 2:1 and 3:1 resonances, with a special focus on stable resonance widths and the expansive unstable chaotic resonance zones. Our findings underscore limitations in semi-analytical approaches (i.e., Gallardo's algorithm) used to assess the influence of MMRs in cislunar space, revealing that the CR3BP model more accurately predicts the region of influence of a specific resonance, based on stable libration widths and chaotic resonance zones. PCR3BP widths reveal a  structure hitherto unseen in other astronomical systems that involve much smaller mass ratios $\mu$ (e.g., the asteroid and Kuiper belts). 
Analysis of unstable resonant periodic orbits has identified regions of chaotic dynamics characterized by interactions between stable and unstable manifolds. 
Specifically, we have quantified the extent of chaotic zones associated with the 2:1 and 3:1 lunar MMRs, and depicted them in geocentric orbital elements. 
Furthermore, by comparing with current cataloged xGEO space objects like TESS, IBEX and Spektr-R, based on TLEs and higher-fidelity ephemeris propagations, we find that our larger regions provide a good fit with objects known to be in the 3:1 and 2:1 MMRs compared to previous approaches. This is particularly true for the IBEX and TESS spacecraft, respectively, which highlights the Earth-Moon system as an astrodynamics laboratory with significant differences compared to other celestial systems. Employing these newly established resonance widths and chaotic zones for a wider range of MMRs --- an atlas of MMRs --- allows for an enhanced determination of whether space assets reside in stable or unstable orbital regimes. 

This examination of MMRs using a global dynamics approach (using the PCR3BP as our model) enhances our understanding of the intricate dynamics of xGEO objects, with implications for mission analysis and design and cislunar SDA and operations. 
Future research directions will focus on refining methodologies to precisely determine resonance widths and chaotic zones, moving beyond qualitative assessments. Additionally, we aim to investigate potential heteroclinic transfers between various resonances and the L1 Lyapunov orbit. Furthermore, we aim to explore the influence of 3-dimensionality (e.g., inclination), a significant factor contributing to the presence of secular resonances.  Further investigation will involve studying chaotic transport across resonances through analysis of lobe dynamics and flux across chaotic regions, as well as extending the study to utilizing 4-dimensional Poincar\'e maps.

\section*{Appendix}

\subsection{Continuous Function in \texorpdfstring{$\ell$}{l} for Perigee Detection}
We introduce a continuous real-valued function for event detection during numerical integration,

\begin{equation}
    h(\ell) = \cos \ell + \tfrac{1}{4} \sin \ell -1,
\end{equation}
$h(\ell)$ only crosses zero in an increasing direction when $\ell=0$ (perigee), effectively preventing false detections of perigee during apogee where $\ell=\pi$. If the spacecraft comes close to the Moon near apogee $\ell=\pi$, then the mean anomaly $\ell$ can momentarily start decreasing near that point. If one uses an events function that, as $\ell$ increases, crosses zero at perigee and apogee in increasing and decreasing directions respectively, this momentary decrease in $\ell$ near apogee can trigger a false detection of $\ell=0$ even when in reality the spacecraft is going through $\ell=\pi$. The above $h$ crosses zero at $\ell = 0$ and $\ell \approx 0.49$ rad in increasing/decreasing directions, respectively; at the latter point the spacecraft is not influenced strongly enough by the Moon for $\ell$ to start decreasing, so the false detection issue is avoided.

\subsection{Transformation from the Inertial Ecliptic frame (ECLIPJ2000) to the Non-Dimensional Synodic Frame} 

In order to plot high-fidelity Horizon and Cowell's 4-body integration data on the Poincar\'e map, a conversion of spacecraft state is required from the inertial geocentric ecliptic frame to the synodic frame under the CR3BP assumptions described in \cite{goddard1989}, Section 3.3.14. For this conversion, the position and velocity vectors of the Moon are also needed first. They are obtained in the inertial geocentric equatorial frame using the MATLAB function \textit{planetEphemeris}, and then are converted into the ecliptic frame using the following rotation matrix, Eq. \ref{eq:obliq}:
\begin{equation}
\label{eq:obliq}
    R_{eci}^{eclip}= 
    \begin{bmatrix}
        1 & 0 & 0\\
        0 & \cos{\varepsilon} & \sin{\varepsilon}\\
        0 &-\sin{\varepsilon} & \cos{\varepsilon}
    \end{bmatrix},     
\end{equation}      
where $\varepsilon$ denotes the obliquity of the ecliptic given by,
\begin{equation}
    \varepsilon = \frac{\pi}{180}\left( 23 + \frac{26}{60} + \frac{21.412}{3600} \right) \,\, \text{rad}.
\end{equation}

Denote the position and velocity of the Moon in the inertial geocentric ecliptic frame as $\mathbf{r}_m$ and  $\mathbf{v}_m$.
The angular velocity of the Moon, assuming circular motion, is then calculated as,
\begin{equation}
    \omega_m = \sqrt{\frac{G(m_e+m_m)}{|\mathbf{r}_m|^3}}.
\end{equation}
The Moon's unit angular momentum vector is given by
\begin{equation}
    \mathbf{\hat{h}} = \frac{\mathbf{r}_m \times \mathbf{v}_m}{|\mathbf{r}_m \times \mathbf{v}_m|}.
\end{equation}
Hence, the Moon's angular velocity vector becomes
\begin{equation}
    \boldsymbol\omega_m = \omega_m \mathbf{\hat{h}},
\end{equation}
and the velocity of the Moon under the circular motion assumption becomes
\begin{equation}
    \mathbf{v}_{m,cir} = \boldsymbol\omega_m \times \mathbf{r}_m.
\end{equation}

The synodic coordinate system moves about the primary (Earth) with the same angular speed as the secondary (the Moon). The instantaneous unit basis vectors $\mathbf{\hat{x}},\mathbf{\hat{y}},\mathbf{\hat{z}}$ for the synodic frame can be defined using the position vector of the Moon from the Earth ($\mathbf{r}_m$) and the velocity vector of the Moon under circular assumption ($\mathbf{v}_m = \mathbf{v}_{m,cir}$) as follows,
\begin{equation}
\begin{matrix}
        \mathbf{\hat{x}} = \frac{\mathbf{r}_m}{|\mathbf{r}_m|}, \\
        \mathbf{\hat{y}} = \frac{(\mathbf{r}_m \times \mathbf{v}_m) \times \mathbf{r}_m}{|(\mathbf{r}_m \times \mathbf{v}_m)| |\mathbf{r}_m|}, \\
        \mathbf{\hat{z}} = \frac{\mathbf{r}_m \times \mathbf{v}_m}{|\mathbf{r}_m \times \mathbf{v}_m|}.
\end{matrix}
\end{equation}

In the absence of external torques (such as perturbations due to other solar system bodies), the rates of change of the synodic frame unit vectors $\mathbf{\hat{x}},\mathbf{\hat{y}},\mathbf{\hat{z}}$ become,
\begin{equation}
\begin{matrix}
        \dot{\hat{\mathbf{x}}} = \frac{\mathbf{v}_m}{|\mathbf{r}_m|} - \frac{\mathbf{r}_m(\mathbf{r}_m \cdot \mathbf{v}_m)}{|\mathbf{r}_m|^3}, \\
        \dot{\hat{\mathbf{y}}} = \frac{(\mathbf{r}_m \times \mathbf{v}_m) \times \mathbf{v}_m}{|\mathbf{r}_m| |\mathbf{r}_m \times \mathbf{v}_m|} - \frac{(\mathbf{r}_m \times \mathbf{v}_m)}{|\mathbf{r}_m|^3|\mathbf{r}_m \times \mathbf{v}_m|} ((\mathbf{r}_m \times \mathbf{v}_m) \times \mathbf{r}_m), \\
        \dot{\hat{\mathbf{z}}} = 0.
\end{matrix}
\end{equation}

The above expressions yield $\mathbf{\hat{x}}, \mathbf{\hat{y}}, \mathbf{\hat{z}}, \dot{\hat{\mathbf{x}}}, \dot{\hat{\mathbf{y}}}, \dot{\hat{\mathbf{z}}}$ in geocentric inertial ecliptic coordinates. Thus, the rotation matrix to convert from geocentric inertial ecliptic frame to synodic frame is as follows \cite{goddard1989},
\begin{equation}
\label{eq:inertial-synodic}
    R_{iner}^{syn} = 
    \begin{bmatrix}
        \Tilde{Q} & \Tilde{0}\\
        \dot{\Tilde{Q}} & \Tilde{Q}\\
    \end{bmatrix},
\end{equation}

where, interpreting $\mathbf{\hat{x}}, \mathbf{\hat{y}}, \mathbf{\hat{z}}$ as $3 \times 1$ column vectors, then $\Tilde{Q}, \dot{\Tilde{Q}} $ are $3 \times 3$ matrices given by,
\begin{equation}
    \Tilde{Q} = \begin{bmatrix}
        \mathbf{\hat{x}} &
        \mathbf{\hat{y}} &
        \mathbf{\hat{z}}
    \end{bmatrix}^T, \hspace{1cm}
    \dot{\Tilde{Q}} = 
    \begin{bmatrix}
        \dot{\hat{\mathbf{x}}} &
        \dot{\hat{\mathbf{y}}} &
        \dot{\hat{\mathbf{z}}}
    \end{bmatrix}^T.
\end{equation}

The mass parameter of the Earth-Moon system can be defined as,
\begin{equation}
    \mu = \frac{m_m}{m_m + m_e}.
\end{equation}

The position and velocity of the barycenter of the synodic frame can be expressed in the geocentric inertial frame as,
\begin{equation}
    \begin{split}
        \mathbf{R}_0 = \mu \mathbf{r}_m, \\
        \mathbf{V}_0 = \mu \mathbf{v}_m. \\
    \end{split}
\end{equation}

Now, to convert spacecraft state from inertial geocentric ecliptic to synodic coordinates, first the origin is shifted from the primary to the barycenter of the system. Then, the inertial position and velocity of spacecraft ($\mathbf{r}$ and $\mathbf{v}$) are transformed into the rotating synodic reference frame using Eq. \ref{eq:inertial-synodic} as,
\begin{equation}
\label{eq:conversion:I2S}
    \begin{bmatrix}
        \mathbf{r}_{syn}\\
        \mathbf{v}_{syn}
    \end{bmatrix} = R_{iner}^{syn}
     \begin{bmatrix}
        \mathbf{r} - \mathbf{R}_0\\
        \mathbf{v} - \mathbf{V}_0\\
    \end{bmatrix}.
\end{equation}

Finally, the vectors $\mathbf{r}_{syn}$ and $\mathbf{v}_{syn}$ obtained in Eq. \ref{eq:conversion:I2S} are non-dimensionalized using the magnitude of instantaneous position vector of the Moon. The non-dimensional constants are given as,
\begin{equation}
    \begin{split}
        l^* = |\mathbf{r}_m|, \\
        \mu^* = G(m_e + m_m), \\
        t^* = \sqrt{ \frac{l^*{}^3}{\mu^*}}.
    \end{split}
\end{equation}

The non-dimensionalized spacecraft state vectors to be used with the CR3BP then become,
\begin{equation}
    \begin{split}
        \mathbf{r}_{syn, non-dim} = \frac{\mathbf{r}_{syn}}{l^*}, \\
        \mathbf{v}_{syn, non-dim} = \mathbf{v}_{syn} * \frac{t^*}{l^*}.
    \end{split}
\end{equation}

\subsection{Integral Invariants of Poincar\'e-Cartan}
All Hamiltonian systems must preserve the Poincar\'e-Cartan integral invariants under a canonical transformation. That is, a set in $2n$-dimensional phase space can be projected onto $n$ 2D planes formed by a conjugate-coordinate axis
system; the oriented sum of the areas of these projections is preserved even as the set evolves under Hamiltonian dynamics. Our 2-dimensional Poincar\'e surface of section at perigee is an oriented projection of a 4-dimensional Hamiltonian system and follows area preservation, as discussed later. In the planar CR3BP model (4-dim. phase space) the canonical conjugate-coordinate pairs are represented by $(g,G)$ and $(\ell,L)$. In this planar case, $g$ represents the same synodic longitude of perigee $\varpi$ defined earlier. Extending to the spatial CR3BP model the phase space becomes 6-dimensional, introducing an additional canonical pair $(h,H)$ and redefining $g$ as the (inertial) \emph{argument} of perigee rather than $\varpi$. These canonical coordinates are called synodic Delaunay variables, and are closely related to orbital elements as follows:

\begin{align}
    \ell =M,\\
    L = \sqrt{(1-\mu) a},\\
    g = \omega,\\
    G = L\sqrt{1-e^2},\\
    h = \Omega,\\
    H = G\cos{i},
\end{align}
where $\Omega$ above is the synodic longitude of ascending node, i.e. the angle between the ascending node direction and the \emph{rotating frame} $x$-axis.

Since we take an oriented surface of section $\Sigma_C$ at perigee $\ell=0$, the area of the projection of any subset of $\Sigma_C$ in the $(\ell,L)$ plane goes to zero. In the PCR3BP, the entire (invariant) area sum is thus projected onto the $(g,G)$ plane. However, in the spatial CR3BP 6-dimensional phase space, the sum of oriented areas projected onto the three canonical coordinate-conjugate pairs --- $(g,G)$, $(h,H)$, and $(\ell,L)$ --- remains invariant through sequential mappings at perigee, i.e., through all resonant islands.

For resonant objects such as TESS and IBEX, perigee mappings within the spatial CR3BP produce closed librating curves in both the $(g,G)$ and $(h,H)$ planes. With the perigee surface of section condition $(\ell=0)$, the area in the $(\ell,L)$ plane again becomes zero. Numerical analysis confirms that the sum of the areas enclosed by librating islands in both the $(g,G)$ and $(h,H)$ planes remains constant across all resonant islands. Since, $\varpi=g+\Omega$, the sum of the areas enclosed by librating islands in both the $(\varpi,G)$ and $(h,H)$ also remains constant across all resonant islands, as qualitatively seen in Fig.\ \ref{fig:map_individual}. Given the close relationship between Delaunay variables and traditional orbital elements, our simulations further reveal that the summed areas in the $(\varpi,a)$ and $(\Omega,i)$ planes also remain approximately constant, an intriguing observation warranting additional research to elucidate the underlying dynamics.

\section*{Acknowledgments}

B.K. was supported in part by the National Science Foundation under award no.\ DMS-2202994, and in part by the US Air Force Office of Scientific Research (AFOSR) under Award No. FA8655-24-1-7012. Part of this research was carried out at the Jet Propulsion Laboratory, California Institute of Technology, under a contract with the National Aeronautics and Space Administration. This research was also supported by funding from the AFOSR under Award No.\ FA9550-24-1-0194.

The present form of this manuscript has greatly benefited from the critical comments and valuable suggestions of numerous colleagues. We are especially grateful to Vladislav Sidorenko of the Keldysh Institute of Applied Mathematics and Renu Malhotra of the Lunar and Planetary Laboratory for insightful discussions on the widths of mean-motion resonances in Solar-System dynamics. A.J.R. would also like to thank Tabar\'e Gallardo of the University of the Republic (Uruguay) for his guidance on the use of his semi-analytical method, and Di Wu of Embry-Riddle Aeronautical University for his early implementation of Gallardo's codes.

{\footnotesize \bibliography{BIBLIOGRAPHY}}

\end{document}